\documentclass[aps,preprint]{revtex4}%
\usepackage{amsfonts}
\usepackage{amsmath}
\usepackage{amssymb}
\usepackage{graphicx}%
\begin{document}
\preprint{ }

\vspace*{1cm}

\begin{center}

{\Large Ground state energy of dilute neutron matter at}

{\Large next-to-leading order in lattice chiral effective field theory}%
\vspace*{0.75cm}

{Evgeny Epelbaum$^{a,b}$, Hermann~Krebs$^{b,a}$, Dean~Lee$^{c,b}$,
Ulf-G.~Mei{\ss }ner$^{b,a}$}\vspace*{0.75cm}

$^{a}$\textit{Institut f\"{u}r Kernphysik (IKP-3) and J\"{u}lich Center for
Hadron Physics,}

\textit{Forschungszentrum J\"{u}lich, D-52425 J\"{u}lich, Germany }

$^{b}$\textit{Helmholtz-Institut f\"{u}r Strahlen- und Kernphysik (Theorie)}

\textit{and Bethe Center for Theoretical Physics, }\linebreak%
\textit{Universit\"{a}t Bonn, D-53115 Bonn, Germany }

$^{c}$\textit{Department of Physics, North Carolina State University, Raleigh,
NC 27695, USA}

\vspace*{0.75cm}

{\large Abstract}
\end{center}

We present lattice calculations for the ground state energy of dilute neutron
matter at next-to-leading order in chiral effective field theory. \ This study
follows a series of recent papers on low-energy nuclear physics using chiral
effective field theory on the lattice. \ In this work we introduce an improved
spin- and isospin-projected leading-order action which allows for a
perturbative treatment of corrections at next-to-leading order and smaller
estimated errors. \ Using auxiliary fields and Euclidean-time projection Monte
Carlo, we compute the ground state of $8$, $12,$ and $16$ neutrons in a
periodic cube, covering a density range from 2\% to 10\% of normal nuclear density.

\pagebreak

\section{Introduction}

Chiral effective field theory for low-energy nucleons on the lattice has been
investigated in several recent papers. \ In Ref.~\cite{Borasoy:2006qn} chiral
effective field theory was considered at leading order (LO) using two
different lattice actions. \ These actions, LO$_{1}$ and LO$_{2}$, each
include the leading-order interactions in Weinberg's power counting scheme
\cite{Weinberg:1990rz,Weinberg:1991um}. \ The difference is that in LO$_{1}$
the nucleon-nucleon \textquotedblleft contact\textquotedblright\ interactions
are point-like while in LO$_{2}$ they are smeared using a Gaussian function.
\ These smeared\ interactions in LO$_{2}$ were used to better reproduce
$S$-wave phase shifts for nucleon momenta up to the pion mass. \ If the
effective field theory expansion is converging properly then low-energy
physical observables computed using LO$_{1}$ and LO$_{2}$ should agree up to
differences of the size of next-to-leading order (NLO) corrections.
\ Similarly when NLO corrections are included, agreement should be comparable
to corrections at next-to-next-to-leading order (NNLO).

In Ref.~\cite{Borasoy:2007vi} the spherical wall method \cite{Borasoy:2007vy}
was used to calculate nucleon-nucleon scattering phase shifts and the $S$-$D$
mixing angle for LO$_{1}$ and LO$_{2}$ at spatial lattice spacing $a=(100$
MeV$)^{-1}$ and temporal lattice spacing $a_{t}=(70$ MeV$)^{-1}$. \ In a
companion paper \cite{Borasoy:2007vk} the same LO$_{2}$ lattice action was
reproduced using auxiliary fields, and the ground state energy of dilute
neutron matter was calculated using projection Monte Carlo at densities
ranging from 2\% to 8\% of normal nuclear density. \ For each Monte Carlo
configuration next-to-leading order corrections were computed using
first-order perturbation theory. \ Simulations using the lattice action
LO$_{1}$ were also attempted, however strong complex phase oscillations
prevented an accurate calculation.

\begin{figure}
[ptb]
\begin{center}
\includegraphics[
height=3.026in,
width=3.5959in
]%
{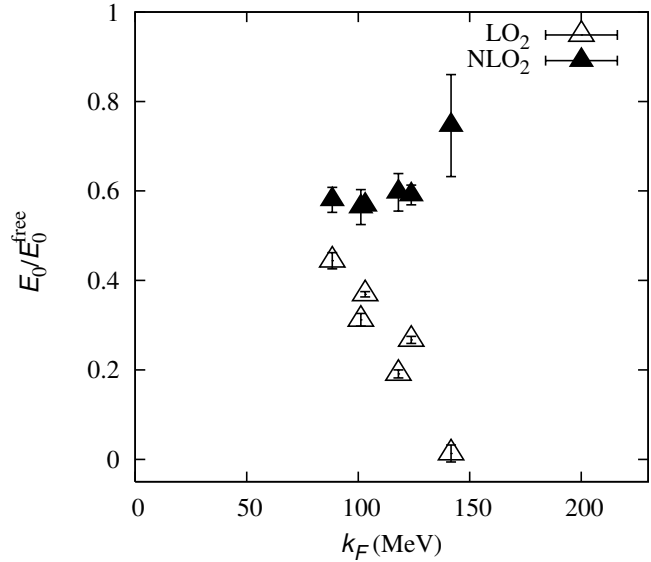}%
\caption{Ground state energy ratio $E_{0}/E_{0}^{\text{free}}$ for dilute
neutron matter versus Fermi momentum $k_{F}$ for LO$_{2}$ and NLO$_{2}$
\cite{Borasoy:2007vk}.}%
\label{kf_xsi_lo2}%
\end{center}
\end{figure}

Ground state energy results using the LO$_{2}$ action at leading order and
next-to-leading order are shown in Fig.~\ref{kf_xsi_lo2}. \ The energy is
plotted as a fraction of the ground state energy for non-interacting neutrons
at the same Fermi momentum $k_{F}$.%
For $k_{F}$ less than $100$ MeV, the difference between results at leading
order and next-to-leading order is small enough that the convergence of the
effective theory appears reliable. \ However for $k_{F}$ greater than $100$
MeV the difference is relatively large, and the perturbative treatment of NLO
corrections seems questionable. \ The analysis in Ref.~\cite{Borasoy:2007vk}
found that much of the difference between the LO$_{2}$ and NLO$_{2}$ results
could be ascribed to differences in the $P$-wave phase shifts. \ Although
helpful in $S$-wave channels, the Gaussian smearing used in LO$_{2}$ produces
unphysical attractive forces in each $P$-wave channel which must be cancelled
at next-to-leading order. \ In this paper we introduce a new leading-order
action LO$_{3}$ that solves this problem. \ The new action equals LO$_{2}$ in
each $S$-wave channel but matches LO$_{1}$ in each $P$-wave channel. \ We
construct the new action using projection operators for the
spin-singlet/isospin-triplet and spin-triplet/isospin-singlet channels.

The paper is organized as follows. \ We first review the effective potential
for chiral effective field theory up to next-to-leading order and
simplifications that can be made at low cutoff momentum. \ We also summarize
the lattice transfer matrix formalism for LO$_{1}$ and LO$_{2}$. The new
action LO$_{3}$ is then introduced and phase shifts and the $S$-$D$ mixing
angle are computed up to next-to-leading order. \ After this we rewrite the
LO$_{3}$ transfer matrix in terms of one-body interactions with auxiliary
fields. \ This allows us to simulate the ground state of the many-neutron
system up to next-to-leading order using projection Monte Carlo. \ We compare
the new results obtained using LO$_{3}$ and NLO$_{3}$ with the LO$_{2}$ and
NLO$_{2}$ results from Ref.~\cite{Borasoy:2007vk} and other published data in
the literature. \ We also analyze the ground state energy ratio $E_{0}%
/E_{0}^{\text{free}}$ as an expansion near the unitarity limit.

\section{Chiral effective field theory}

\subsection{Effective potential}

In our notation $\vec{q}$ denotes the $t$-channel momentum transfer for
nucleon-nucleon scattering and $\vec{k}$ is the $u$-channel exchanged momentum
transfer. At leading order in the Weinberg power-counting scheme
\cite{Weinberg:1990rz,Weinberg:1991um} the nucleon-nucleon effective potential
includes two independent contact terms and instantaneous one-pion exchange
(OPEP),%
\begin{equation}
V_{\text{LO}}=V^{(0)}+V^{\text{OPEP}}, \label{VLO}%
\end{equation}%
\begin{equation}
V^{(0)}=C_{S}+C_{T}\left(  \vec{\sigma}_{1}\cdot\vec{\sigma}_{2}\right)  ,
\label{V0}%
\end{equation}%
\begin{equation}
V^{\text{OPEP}}=-\left(  \frac{g_{A}}{2f_{\pi}}\right)  ^{2}\boldsymbol\tau
_{1}\cdot\boldsymbol\tau_{2}\frac{\left(  \vec{\sigma}_{1}\cdot\vec{q}\right)
\left(  \vec{\sigma}_{2}\cdot\vec{q}\right)  }{q^{\,2}+m_{\pi}^{2}}.
\label{VOPEP}%
\end{equation}
The vector arrow in $\vec{\sigma}$ signifies the three-vector index for spin,
and the boldface for $\boldsymbol\tau$ signifies the three-vector index for
isospin. \ For physical constants we take $m=938.92$ MeV as the nucleon mass,
$m_{\pi}=138.08$ MeV as the pion mass, $f_{\pi}=93$ MeV as the pion decay
constant, and $g_{A}=1.26$ as the nucleon axial charge.

At next-to-leading order the effective potential has seven independent contact
terms carrying two powers of momentum, corrections to the two LO\ contact
terms, and instantaneous two-pion exchange (TPEP)
\cite{Ordonez:1992xp,Ordonez:1993tn,Ordonez:1996rz,Epelbaum:1998ka,Epelbaum:1999dj}%
. \ Following the notation of Ref.~\cite{Epelbaum:1998ka,Epelbaum:1999dj} we
have%
\begin{equation}
V_{\text{NLO}}=V_{\text{LO}}+\Delta V^{(0)}+V^{(2)}+V_{\text{NLO}%
}^{\text{TPEP}}. \label{VNLO}%
\end{equation}
The NLO contact interactions are given by%
\begin{equation}
\Delta V^{(0)}=\Delta C_{S}+\Delta C_{T}\left(  \vec{\sigma}_{1}\cdot
\vec{\sigma}_{2}\right)  , \label{dV0}%
\end{equation}%
\begin{align}
V^{(2)}  &  =C_{1}q^{2}+C_{2}k^{2}+\left(  C_{3}q^{2}+C_{4}k^{2}\right)
\left(  \vec{\sigma}_{1}\cdot\vec{\sigma}_{2}\right)  +iC_{5}\frac{1}%
{2}\left(  \vec{\sigma}_{1}+\vec{\sigma}_{2}\right)  \cdot\left(  \vec
{q}\times\vec{k}\right) \nonumber\\
&  +C_{6}\left(  \vec{\sigma}_{1}\cdot\vec{q}\right)  \left(  \vec{\sigma}%
_{2}\cdot\vec{q}\right)  +C_{7}\left(  \vec{\sigma}_{1}\cdot\vec{k}\right)
\left(  \vec{\sigma}_{2}\cdot\vec{k}\right)  , \label{V2}%
\end{align}
and the NLO two-pion exchange potential is \cite{Friar:1994,Kaiser:1997mw}%
\begin{align}
V_{\text{NLO}}^{\text{TPEP}}  &  =-\frac{\boldsymbol\tau_{1}\cdot
\boldsymbol\tau_{2}}{384\pi^{2}f_{\pi}^{4}}L(q)\left[  4m_{\pi}^{2}\left(
5g_{A}^{4}-4g_{A}^{2}-1\right)  +q^{2}\left(  23g_{A}^{4}-10g_{A}%
^{2}-1\right)  +\frac{48g_{A}^{4}m_{\pi}^{4}}{4m_{\pi}^{2}+q^{2}}\right]
\nonumber\\
&  -\frac{3g_{A}^{4}}{64\pi^{2}f_{\pi}^{4}}L(q)\left[  \left(  \vec{q}%
\cdot\vec{\sigma}_{1}\right)  \left(  \vec{q}\cdot\vec{\sigma}_{2}\right)
-q^{2}\left(  \vec{\sigma}_{1}\cdot\vec{\sigma}_{2}\right)  \right]  ,
\label{VTPEPNLO}%
\end{align}
where%
\begin{equation}
L(q)=\frac{1}{2q}\sqrt{4m_{\pi}^{2}+q^{2}}\ln\frac{\sqrt{4m_{\pi}^{2}+q^{2}%
}+q}{\sqrt{4m_{\pi}^{2}+q^{2}}-q}. \label{Lq}%
\end{equation}
Recent reviews of chiral effective field theory can be found in
Ref.~\cite{vanKolck:1999mw,Bedaque:2002mn,Epelbaum:2005pn,Epelbaum:2008ga}.

\subsection{Power counting and cutoff momentum}

There have been a number of studies on the short-distance behavior of the
one-pion exchange potential and the consistency of the Weinberg power-counting
scheme. \ An alternative scheme known as KSW power counting was proposed
\cite{Kaplan:1996xu,Kaplan:1998tg,Kaplan:1998we}. \ This scheme is based on a
perturbative treatment of the one-pion exchange potential and allows for
systematic control of ultraviolet divergences in the effective theory.
\ Unfortunately convergence at higher orders was found to be poor in some
partial waves for momenta comparable to the pion mass \cite{Fleming:1999ee}.

Other power counting alternatives have also been proposed. \ In one scheme the
leading $f_{\pi}^{-2}r^{-3}$ short-distance singularity is treated
non-perturbatively while the remainder of the one-pion potential is introduced
as a perturbative expansion in powers of $m_{\pi}$ \cite{Beane:2001bc}. \ More
recently a different power counting modification was proposed in which
one-pion exchange is treated non-perturbatively in lower angular momentum
channels along with higher-derivative counterterms promoted to leading order
\cite{Nogga:2005hy}. \ Further investigations of this approach in higher
partial waves and power counting with one-pion exchange were considered in
Ref.~\cite{Birse:2005um,Birse:2007sx}.

On the lattice the ultraviolet momentum cutoff is inversely proportional to
the lattice spacing, $\Lambda=\pi$/$a$. \ For simple calculations of
two-nucleon scattering on the lattice we could take any lattice spacing
satisfying $\Lambda\gg m_{\pi}$. \ However in few- and many-nucleon
calculations where we use Euclidean-time projection and auxiliary-field Monte
Carlo methods, severe numerical problems appear when $\Lambda$ is very large.
\ In some attractive channels the problem is due to spurious deeply-bound
states which may appear at sufficiently large $\Lambda$. \ In other channels
one faces the problem due to short-range hard-core repulsion at very large
$\Lambda$. \ This is manifested as sign or complex phase oscillations which
scale exponentially with system size and strength of the repulsive interaction.

To avoid these problems we consider lattice simulations where the cutoff
momentum is only a few times the pion mass. \ In this study we take
$\Lambda=314$ MeV $\approx2.3m_{\pi}$, corresponding with $a^{-1}=100$ MeV.
\ For this low cutoff scale the advantages of the alternative power-counting
schemes discussed above are numerically insignificant \cite{Epelbaum:2006pt},
and so we use standard Weinberg power counting. \ For nearly all $\left\vert
q\right\vert <\Lambda$ we can expand the two-pion exchange potential in powers
of $q^{2}/(4m_{\pi}^{2}),$%
\begin{equation}
L(q)=1+\frac{1}{3}\frac{q^{2}}{4m_{\pi}^{2}}+\cdots,
\end{equation}%
\begin{equation}
\frac{4m_{\pi}^{2}}{4m_{\pi}^{2}+q^{2}}L(q)=1-\frac{2}{3}\frac{q^{2}}{4m_{\pi
}^{2}}+\cdots,
\end{equation}%
\begin{align}
V_{\text{NLO}}^{\text{TPEP}}  &  =-\frac{\boldsymbol\tau_{1}\cdot
\boldsymbol\tau_{2}}{384\pi^{2}f_{\pi}^{4}}\left[  4m_{\pi}^{2}\left(
8g_{A}^{4}-4g_{A}^{2}-1\right)  +\frac{2}{3}q^{2}\left(  34g_{A}^{4}%
-17g_{A}^{2}-2\right)  +O\left(  \left(  \tfrac{q^{2}}{4m_{\pi}^{2}}\right)
^{2}\right)  \right] \nonumber\\
&  -\frac{3g_{A}^{4}}{64\pi^{2}f_{\pi}^{4}}\left[  \left(  \vec{q}\cdot
\vec{\sigma}_{1}\right)  \left(  \vec{q}\cdot\vec{\sigma}_{2}\right)
-q^{2}\left(  \vec{\sigma}_{1}\cdot\vec{\sigma}_{2}\right)  \right]  \left[
1+O\left(  \tfrac{q^{2}}{4m_{\pi}^{2}}\right)  \right]  . \label{localTPEP}%
\end{align}
This expansion fails to converge only for values of $q$ near the cutoff scale
$\Lambda$ $\approx2.3m_{\pi}$, where the effective theory is already
problematic due to large cutoff effects of size $O\left(  q^{2}/\Lambda
^{2}\right)  $. \ There is no reason to keep the full non-local structure of
$V_{\text{NLO}}^{\text{TPEP}}$ at this lattice spacing. \ Instead we simply
use%
\begin{equation}
V_{\text{LO}}=V^{(0)}+V^{\text{OPEP}},
\end{equation}%
\begin{equation}
V_{\text{NLO}}=V_{\text{LO}}+\Delta V^{(0)}+V^{(2)},
\end{equation}
where the terms in Eq.~(\ref{localTPEP}) with up to two powers of $q$ are
absorbed as a redefinition of the coefficients $\Delta V^{(0)}$ and $V^{(2)}$.
\ This same approach can be applied to the two-pion exchange potential at
next-to-next-to-leading order and higher-order $n$-pion exchange potentials.

\section{Lattice formalism}

\subsection{Lattice notation}

In this paper we assume exact isospin symmetry and neglect electromagnetic
interactions. \ We use $\vec{n}$ to represent integer-valued lattice vectors
on a three-dimensional spatial lattice and either $\vec{p},$ $\vec{q}$, or
$\vec{k}$ to represent integer-valued momentum lattice vectors.$\ \ \hat
{l}=\hat{1}$, $\hat{2}$, $\hat{3}$ are unit lattice vectors in the spatial
directions, $a$ is the spatial lattice spacing, and $L$ is the length of the
cubic spatial lattice in each direction. \ We use the Euclidean transfer
matrix formalism defined in \cite{Borasoy:2006qn} with lattice time step
$a_{t}$, and the integer $n_{t}$ labels the time steps. \ We define
$\alpha_{t}$ as the ratio between lattice spacings, $\alpha_{t}=a_{t}/a$.
\ Throughout we use dimensionless parameters and operators, which correspond
with physical values multiplied by the appropriate power of $a$. \ Final
results are presented in physical units with the corresponding unit stated
explicitly. \ As in \cite{Borasoy:2006qn} the spatial lattice spacing is
$a=(100$ MeV$)^{-1}$ and temporal lattice spacing is $a_{t}=(70$ MeV$)^{-1}$.

We use $a$ and $a^{\dagger}$ to denote annihilation and creation operators.
\ To avoid confusion we make explicit in our lattice notation all spin and
isospin indices using%
\begin{align}
a_{0,0}  &  =a_{\uparrow,p},\text{ \ }a_{0,1}=a_{\uparrow,n},\\
a_{1,0}  &  =a_{\downarrow,p},\text{ \ }a_{1,1}=a_{\downarrow,n}.
\end{align}
The first subscript is for spin and the second subscript is for isospin. \ We
use $\tau_{I}$ with $I=1,2,3$ to represent Pauli matrices acting in isospin
space and $\sigma_{S}$ with $S=1,2,3$ to represent Pauli matrices acting in
spin space. \ We also use the letters $S$ and $I$ to denote the total spin and
total isospin for the two-nucleon system. \ The intended meaning in each case
should be clear from the context. \ We use the eight vertices of a unit cube
on the lattice to define spatial derivatives. \ For each spatial direction
$l=1,2,3$ and any lattice function $f(\vec{n})$, let%
\begin{equation}
\Delta_{l}f(\vec{n})=\frac{1}{4}\sum_{\substack{\nu_{1},\nu_{2},\nu_{3}%
=0,1}}(-1)^{\nu_{l}+1}f(\vec{n}+\vec{\nu}),\qquad\vec{\nu}=\nu_{1}\hat{1}%
+\nu_{2}\hat{2}+\nu_{3}\hat{3}. \label{derivative}%
\end{equation}
We also define the double spatial derivative along direction $l$,%
\begin{equation}
\triangledown_{l}^{2}f(\vec{n})=f(\vec{n}+\hat{l})+f(\vec{n}-\hat{l}%
)-2f(\vec{n}).
\end{equation}

\subsection{Densities and current densities}

We define the local density of nucleons at lattice site $\vec{n}$,%
\begin{equation}
\rho^{a^{\dagger},a}(\vec{n})=\sum_{i,j=0,1}a_{i,j}^{\dagger}(\vec{n}%
)a_{i,j}(\vec{n}).
\end{equation}
This is invariant under Wigner's SU(4) symmetry transforming all spin and
isospin degrees of freedom \cite{Wigner:1937}. \ Similarly we define a local
spin density for $S=1,2,3,$%
\begin{equation}
\rho_{S}^{a^{\dagger},a}(\vec{n})=\sum_{i,j,i^{\prime}=0,1}a_{i,j}^{\dagger
}(\vec{n})\left[  \sigma_{S}\right]  _{ii^{\prime}}a_{i^{\prime},j}(\vec{n}),
\end{equation}
isospin density for $I=1,2,3,$%
\begin{equation}
\rho_{I}^{a^{\dagger},a}(\vec{n})=\sum_{i,j,j^{\prime}=0,1}a_{i,j}^{\dagger
}(\vec{n})\left[  \tau_{I}\right]  _{jj^{\prime}}a_{i,j^{\prime}}(\vec{n}),
\end{equation}
and spin-isospin density for $S,I=1,2,3,$%
\begin{equation}
\rho_{S,I}^{a^{\dagger},a}(\vec{n})=\sum_{i,j,i^{\prime},j^{\prime}%
=0,1}a_{i,j}^{\dagger}(\vec{n})\left[  \sigma_{S}\right]  _{ii^{\prime}%
}\left[  \tau_{I}\right]  _{jj^{\prime}}a_{i^{\prime},j^{\prime}}(\vec{n}).
\end{equation}

For each static density we also have an associated current density. \ Similar
to the definition of the lattice derivative $\Delta_{l}$ in
Eq.~(\ref{derivative}), we use the eight vertices of a unit cube,
\begin{equation}
\vec{\nu}=\nu_{1}\hat{1}+\nu_{2}\hat{2}+\nu_{3}\hat{3},
\end{equation}
for $\nu_{1},\nu_{2},\nu_{3}=0,1$. \ Let $\vec{\nu}(-l)$ for $l=1,2,3$ be the
result of reflecting the $l^{\text{th}}$-component of $\vec{\nu}$ about the
center of the cube,%
\begin{equation}
\vec{\nu}(-l)=\vec{\nu}+(1-2\nu_{l})\hat{l}.
\end{equation}
Omitting factors of $i$ and $1/m$, we can write the $l^{\text{th}}$-component
of the SU(4)-invariant current density as%
\begin{equation}
\Pi_{l}^{a^{\dagger},a}(\vec{n})=\frac{1}{4}\sum_{\substack{\nu_{1},\nu
_{2},\nu_{3}=0,1}}\sum_{i,j=0,1}(-1)^{\nu_{l}+1}a_{i,j}^{\dagger}(\vec{n}%
+\vec{\nu}(-l))a_{i,j}(\vec{n}+\vec{\nu}).
\end{equation}
Similarly the $l^{\text{th}}$-component of spin current density is%
\begin{equation}
\Pi_{l,S}^{a^{\dagger},a}(\vec{n})=\frac{1}{4}\sum_{\substack{\nu_{1},\nu
_{2},\nu_{3}=0,1}}\sum_{i,j,i^{\prime}=0,1}(-1)^{\nu_{l}+1}a_{i,j}^{\dagger
}(\vec{n}+\vec{\nu}(-l))\left[  \sigma_{S}\right]  _{ii^{\prime}}a_{i^{\prime
},j}(\vec{n}+\vec{\nu}),
\end{equation}
$l^{\text{th}}$-component of isospin current density is%
\begin{equation}
\Pi_{l,I}^{a^{\dagger},a}(\vec{n})=\frac{1}{4}\sum_{\substack{\nu_{1},\nu
_{2},\nu_{3}=0,1}}\sum_{i,j,j^{\prime}=0,1}(-1)^{\nu_{l}+1}a_{i,j}^{\dagger
}(\vec{n}+\vec{\nu}(-l))\left[  \tau_{I}\right]  _{jj^{\prime}}a_{i,j^{\prime
}}(\vec{n}+\vec{\nu}),
\end{equation}
and $l^{\text{th}}$-component of spin-isospin current density is%
\begin{equation}
\Pi_{l,S,I}^{a^{\dagger},a}(\vec{n})=\frac{1}{4}\sum_{\substack{\nu_{1}%
,\nu_{2},\nu_{3}=0,1}}\sum_{i,j,i^{\prime},j^{\prime}=0,1}(-1)^{\nu_{l}%
+1}a_{i,j}^{\dagger}(\vec{n}+\vec{\nu}(-l))\left[  \sigma_{S}\right]
_{ii^{\prime}}\left[  \tau_{I}\right]  _{jj^{\prime}}a_{i^{\prime},j^{\prime}%
}(\vec{n}+\vec{\nu}).
\end{equation}

\section{Lattice actions}

\subsection{Instantaneous free pion action}

The lattice action for free pions with purely instantaneous propagation is%
\begin{equation}
S_{\pi\pi}(\pi_{I})=\alpha_{t}(\tfrac{m_{\pi}^{2}}{2}+3)\sum_{\vec{n},n_{t}%
,I}\pi_{I}(\vec{n},n_{t})\pi_{I}(\vec{n},n_{t})-\alpha_{t}\sum_{\vec{n}%
,n_{t},I,l}\pi_{I}(\vec{n},n_{t})\pi_{I}(\vec{n}+\hat{l},n_{t}),
\label{pionaction}%
\end{equation}
where $\pi_{I}$ is the pion field labelled with isospin index $I$. \ We note
that pion fields at different time steps $n_{t}$ and $n_{t}^{\prime}$ are
decoupled due to the omission of time derivatives. \ This decoupling among
different time steps generates instantaneous propagation\ in one-pion exchange
diagrams and eliminates radiative pion loops. \ It is convenient to define a
rescaled pion field, $\pi_{I}^{\prime}$,%
\begin{equation}
\pi_{I}^{\prime}(\vec{n},n_{t})=\sqrt{q_{\pi}}\pi_{I}(\vec{n},n_{t}),
\end{equation}%
\begin{equation}
q_{\pi}=\alpha_{t}(m_{\pi}^{2}+6).
\end{equation}
Then%
\begin{equation}
S_{\pi\pi}(\pi_{I}^{\prime})=\frac{1}{2}\sum_{\vec{n},n_{t},I}\pi_{I}^{\prime
}(\vec{n},n_{t})\pi_{I}^{\prime}(\vec{n},n_{t})-\frac{\alpha_{t}}{q_{\pi}}%
\sum_{\vec{n},n_{t},I,l}\pi_{I}^{\prime}(\vec{n},n_{t})\pi_{I}^{\prime}%
(\vec{n}+\hat{l},n_{t}).
\end{equation}

In momentum space the action is%
\begin{equation}
S_{\pi\pi}(\pi_{I}^{\prime})=\frac{1}{L^{3}}\sum_{I,\vec{k}}\pi_{I}^{\prime
}(-\vec{k},n_{t})\pi_{I}^{\prime}(\vec{k},n_{t})\left[  \frac{1}{2}%
-\frac{\alpha_{t}}{q_{\pi}}\sum_{l}\cos\left(  \tfrac{2\pi k_{l}}{L}\right)
\right]  .
\end{equation}
The instantaneous pion correlation function at spatial separation $\vec{n}$ is%
\begin{align}
\left\langle \pi_{I}^{\prime}(\vec{n},n_{t})\pi_{I}^{\prime}(\vec{0}%
,n_{t})\right\rangle  &  =\frac{\int D\pi_{I}^{\prime}\;\pi_{I}^{\prime}%
(\vec{n},n_{t})\pi_{I}^{\prime}(\vec{0},n_{t})\;\exp\left[  -S_{\pi\pi
}\right]  }{\int D\pi_{I}^{\prime}\;\exp\left[  -S_{\pi\pi}\right]  }\text{
\ (no sum on }I\text{)}\nonumber\\
&  =\frac{1}{L^{3}}\sum_{\vec{k}}e^{-i\frac{2\pi}{L}\vec{k}\cdot\vec{n}}%
D_{\pi}(\vec{k}),
\end{align}
where%
\begin{equation}
D_{\pi}(\vec{k})=\frac{1}{1-\tfrac{2\alpha_{t}}{q_{\pi}}\sum_{l}\cos\left(
\tfrac{2\pi k_{l}}{L}\right)  }.
\end{equation}

\subsection{Transfer matrices for LO$_{1}$ and LO$_{2}$}

Roughly speaking, the Euclidean-time transfer matrix is the exponential of the
Hamiltonian, $\exp(-H\Delta t)$, where $\Delta t$ equals one temporal lattice
spacing. \ The normal-ordered transfer matrix for non-interacting nucleons is%
\begin{equation}
M_{\text{free}}=\colon\exp\left(  -H_{\text{free}}\alpha_{t}\right)  \colon,
\end{equation}
where the $::$ symbols indicate normal ordering. \ We use the $O(a^{4}%
)$-improved free lattice Hamiltonian,%
\begin{align}
H_{\text{free}}  &  =\frac{49}{12m}\sum_{\vec{n}}\sum_{i,j=0,1}a_{i,j}%
^{\dagger}(\vec{n})a_{i,j}(\vec{n})\nonumber\\
&  -\frac{3}{4m}\sum_{\vec{n}}\sum_{i,j=0,1}\sum_{l=1,2,3}\left[
a_{i,j}^{\dagger}(\vec{n})a_{i,j}(\vec{n}+\hat{l})+a_{i,j}^{\dagger}(\vec
{n})a_{i,j}(\vec{n}-\hat{l})\right] \nonumber\\
&  +\frac{3}{40m}\sum_{\vec{n}}\sum_{i,j=0,1}\sum_{l=1,2,3}\left[
a_{i,j}^{\dagger}(\vec{n})a_{i,j}(\vec{n}+2\hat{l})+a_{i,j}^{\dagger}(\vec
{n})a_{i,j}(\vec{n}-2\hat{l})\right] \nonumber\\
&  -\frac{1}{180m}\sum_{\vec{n}}\sum_{i,j=0,1}\sum_{l=1,2,3}\left[
a_{i,j}^{\dagger}(\vec{n})a_{i,j}(\vec{n}+3\hat{l})+a_{i,j}^{\dagger}(\vec
{n})a_{i,j}(\vec{n}-3\hat{l})\right]  .
\end{align}

Let us define the two-derivative pion correlator,%
\begin{align}
G_{S_{1}S_{2}}(\vec{n})  &  =\left\langle \Delta_{S_{1}}\pi_{I}^{\prime}%
(\vec{n},n_{t})\Delta_{S_{2}}\pi_{I}^{\prime}(\vec{0},n_{t})\right\rangle
\text{ \ (no sum on }I\text{)}\nonumber\\
&  =\frac{1}{16}\sum_{\nu_{1},\nu_{2},\nu_{3}=0,1}\sum_{\nu_{1}^{\prime}%
,\nu_{2}^{\prime},\nu_{3}^{\prime}=0,1}(-1)^{\nu_{S_{1}}}(-1)^{\nu_{S_{2}%
}^{\prime}}\left\langle \pi_{I}^{\prime}(\vec{n}+\vec{\nu}-\vec{\nu}^{\prime
},n_{t})\pi_{I}^{\prime}(\vec{0},n_{t})\right\rangle \text{.}%
\end{align}
With interactions included, the lattice transfer matrix LO$_{1}$ has the form%
\begin{align}
M_{\text{LO}_{1}}  &  =\colon\exp\left\{  -H_{\text{free}}\alpha_{t}-\frac
{1}{2}C\alpha_{t}\sum_{\vec{n}}\left[  \rho^{a^{\dag},a}(\vec{n})\right]
^{2}-\frac{1}{2}C_{I^{2}}\alpha_{t}\sum_{I}\sum_{\vec{n}}\left[  \rho
_{I}^{a^{\dag},a}(\vec{n})\right]  ^{2}\right. \nonumber\\
&  \qquad+\left.  \frac{g_{A}^{2}\alpha_{t}^{2}}{8f_{\pi}^{2}q_{\pi}}%
\sum_{\substack{S_{1},S_{2},I}}\sum_{\vec{n}_{1},\vec{n}_{2}}G_{S_{1}S_{2}%
}(\vec{n}_{1}-\vec{n}_{2})\rho_{S_{1},I}^{a^{\dag},a}(\vec{n}_{1})\rho
_{S_{2},I}^{a^{\dag},a}(\vec{n}_{2})\right\}  \colon,
\end{align}
where $C$ is the coefficient of the Wigner SU(4)-invariant contact interaction
and $C_{I^{2}}$ is the coefficient of the isospin-dependent contact
interaction. \ For the $S$-wave there are two independent channels
corresponding with the spin-singlet/isospin-triplet and the
spin-triplet/isospin-singlet. \ To reproduce the physical scattering lengths
in each channel we set $C_{S=0,I=1}=-5.021\times10^{-5}$ MeV$^{-2}$ and
$C_{S=1,I=0}=-5.714\times10^{-5}$ MeV$^{-2}$ and use the relations
\begin{equation}
C=\left(  3C_{S=0,I=1}+C_{S=1,I=0}\right)  /4, \label{C_coeff}%
\end{equation}%
\begin{equation}
C_{I^{2}}=\left(  C_{S=0,I=1}-C_{S=1,I=0}\right)  /4. \label{C_I2_coeff}%
\end{equation}
\ 

The LO$_{2}$ transfer matrix is \cite{Borasoy:2006qn}%
\begin{align}
M_{\text{LO}_{2}}  &  =\colon\exp\left\{  -H_{\text{free}}\alpha_{t}%
-\frac{\alpha_{t}}{2L^{3}}\sum_{\vec{q}}f(\vec{q})\left[  C\rho^{a^{\dag}%
,a}(\vec{q})\rho^{a^{\dag},a}(-\vec{q})+C_{I^{2}}\sum_{I}\rho_{I}^{a^{\dag}%
,a}(\vec{q})\rho_{I}^{a^{\dag},a}(-\vec{q})\right]  \right. \nonumber\\
&  \qquad+\left.  \frac{g_{A}^{2}\alpha_{t}^{2}}{8f_{\pi}^{2}q_{\pi}}%
\sum_{\substack{S_{1},S_{2},I}}\sum_{\vec{n}_{1},\vec{n}_{2}}G_{S_{1}S_{2}%
}(\vec{n}_{1}-\vec{n}_{2})\rho_{S_{1},I}^{a^{\dag},a}(\vec{n}_{1})\rho
_{S_{2},I}^{a^{\dag},a}(\vec{n}_{2})\right\}  \colon,
\end{align}
where the momentum-dependent coefficient function $f(\vec{q})$ is defined as
\begin{equation}
f(\vec{q})=f_{0}^{-1}\exp\left[  -b%
{\displaystyle\sum\limits_{l}}
\left(  1-\cos q_{l}\right)  \right]  , \label{fq2}%
\end{equation}
and the normalization factor $f_{0}$ is determined by the condition%
\begin{equation}
f_{0}=\frac{1}{L^{3}}\sum_{\vec{q}}\exp\left[  -b%
{\displaystyle\sum\limits_{l}}
\left(  1-\cos q_{l}\right)  \right]  . \label{f_0}%
\end{equation}
As in Ref.~\cite{Borasoy:2006qn} we use the value $b=0.6$. \ This gives
approximately the correct average effective range for the two $S$-wave
channels when $C$ and $C_{I^{2}}$ are tuned to produce the physical $S$-wave
scattering lengths. \ We set $C_{S=0,I=1}=-3.414\times10^{-5}$ MeV$^{-2}$ and
$C_{S=1,I=0}=-4.780\times10^{-5}$ MeV$^{-2}$ and use the same relations
Eq.~(\ref{C_coeff}) and (\ref{C_I2_coeff}). \ The momentum-dependent function
$f(\vec{q})$ produces the Gaussian-smeared \textquotedblleft
contact\textquotedblright\ interactions discussed in the introduction.

The replacement of pointlike interactions in LO$_{1}$ with Gaussian-smeared
interactions in LO$_{2}$ is similar to the lattice improvement program of
Symanzik used in lattice QCD actions \cite{Symanzik:1983dc,Symanzik:1983gh}.
\ There is a conceptual difference however since we are dealing with an
effective field theory rather than a renormalizable field theory. \ The
higher-order operators we consider do not only cancel lattice artifacts but
also include higher-order interactions of the effective theory. \ In our
lattice calculations the improved\ leading-order\ action is treated
non-perturbatively while higher-order interactions are included as a
perturbative expansion. \ The choice of improved action sets a dividing line
between perturbative and non-perturbative interactions. \ This dividing line
should be immaterial so long as the perturbative expansion converges. \ At any
given order, lattice calculations using different improved actions should
agree up to corrections the size of terms at next order.

\subsection{Transfer matrix for LO$_{3}$}

The Gaussian smearing used in LO$_{2}$ is useful in $S$-wave channels but
produces unphysical attractive forces in $P$-wave channels. \ To avoid this
problem we introduce a new leading-order action LO$_{3}$ that equals LO$_{2}$
in each $S$-wave channel but matches LO$_{1}$ in each $P$-wave channel. \ We
multiply the Gaussian-smeared \textquotedblleft contact\textquotedblright%
\ interactions with projection operators for the spin-singlet/isospin-triplet
channel, $P_{S=0,I=1}$, and the spin-triplet/isospin-singlet channel,
$P_{S=1,I=0}$. \ If we assign labels to the two nucleons, $A$ and $B$, these
projection operators are%
\begin{equation}
P_{S=0,I=1}=\left(  \frac{1}{4}-\frac{1}{4}\sum_{S}\sigma_{S}^{A}\sigma
_{S}^{B}\right)  \left(  \frac{3}{4}+\frac{1}{4}\sum_{I}\tau_{I}^{A}\tau
_{I}^{B}\right)  ,
\end{equation}%
\begin{equation}
P_{S=1,I=0}=\left(  \frac{3}{4}+\frac{1}{4}\sum_{S}\sigma_{S}^{A}\sigma
_{S}^{B}\right)  \left(  \frac{1}{4}-\frac{1}{4}\sum_{I}\tau_{I}^{A}\tau
_{I}^{B}\right)  .
\end{equation}
We can define corresponding momentum-dependent density correlations,%
\begin{align}
V_{S=0,I=1}(\vec{q})  &  =\frac{3}{32}:\rho^{a^{\dag},a}(\vec{q})\rho
^{a^{\dag},a}(-\vec{q}):-\frac{3}{32}:\sum_{S}\rho_{S}^{a^{\dag},a}(\vec
{q})\rho_{S}^{a^{\dag},a}(-\vec{q}):\nonumber\\
&  +\frac{1}{32}:\sum_{I}\rho_{I}^{a^{\dag},a}(\vec{q})\rho_{I}^{a^{\dag}%
,a}(-\vec{q}):-\frac{1}{32}:\sum_{S,I}\rho_{S,I}^{a^{\dag},a}(\vec{q}%
)\rho_{S,I}^{a^{\dag},a}(-\vec{q}):,
\end{align}%
\begin{align}
V_{S=1,I=0}(\vec{q})  &  =\frac{3}{32}:\rho^{a^{\dag},a}(\vec{q})\rho
^{a^{\dag},a}(-\vec{q}):+\frac{1}{32}:\sum_{S}\rho_{S}^{a^{\dag},a}(\vec
{q})\rho_{S}^{a^{\dag},a}(-\vec{q}):\nonumber\\
&  -\frac{3}{32}:\sum_{I}\rho_{I}^{a^{\dag},a}(\vec{q})\rho_{I}^{a^{\dag}%
,a}(-\vec{q}):-\frac{1}{32}:\sum_{S,I}\rho_{S,I}^{a^{\dag},a}(\vec{q}%
)\rho_{S,I}^{a^{\dag},a}(-\vec{q}):.
\end{align}
We use $V_{S=0,I=1}$ and $V_{S=1,I=0}$ to write the leading-order transfer
matrix for LO$_{3}$,%
\begin{align}
M_{\text{LO}_{3}}  &  =\colon\exp\left\{  -H_{\text{free}}\alpha_{t}%
-\frac{\alpha_{t}}{L^{3}}\sum_{\vec{q}}f(\vec{q})\left[  C_{S=0,I=1}%
V_{S=0,I=1}(\vec{q})+C_{S=1,I=0}V_{S=1,I=0}(\vec{q})\right]  \right.
\nonumber\\
&  \qquad+\left.  \frac{g_{A}^{2}\alpha_{t}^{2}}{8f_{\pi}^{2}q_{\pi}}%
\sum_{\substack{S_{1},S_{2},I}}\sum_{\vec{n}_{1},\vec{n}_{2}}G_{S_{1}S_{2}%
}(\vec{n}_{1}-\vec{n}_{2})\rho_{S_{1},I}^{a^{\dag},a}(\vec{n}_{1})\rho
_{S_{2},I}^{a^{\dag},a}(\vec{n}_{2})\right\}  \colon.
\end{align}
The momentum-dependent coefficient function $f(\vec{q})$ is the same as
defined in Eq.~\ref{fq2} and \ref{f_0}.

\subsection{Lattice interactions at next-to-leading-order}

The lattice interactions at next-to-leading order were discussed in
Ref.~\cite{Borasoy:2007vi}. \ We follow the same formalism here. \ We start
with the corrections to the leading-order \textquotedblleft
contact\textquotedblright\ interactions. \ These NLO interactions are chosen
to be point-like rather than smeared operators, and we write the interactions
in the same manner as in Ref.~\cite{Borasoy:2007vi},%
\begin{equation}
\Delta V=\frac{1}{2}\Delta C:\sum\limits_{\vec{n}}\rho^{a^{\dagger},a}(\vec
{n})\rho^{a^{\dagger},a}(\vec{n}):,
\end{equation}%
\begin{equation}
\Delta V_{I^{2}}=\frac{1}{2}\Delta C_{I^{2}}:\sum\limits_{\vec{n},I}\rho
_{I}^{a^{\dagger},a}(\vec{n})\rho_{I}^{a^{\dagger},a}(\vec{n}):.
\end{equation}
At next-to-leading order there are seven independent contact interactions with
two derivatives. \ The basis we choose is%
\begin{equation}
V_{q^{2}}=-\frac{1}{2}C_{q^{2}}:\sum\limits_{\vec{n},l}\rho^{a^{\dagger}%
,a}(\vec{n})\triangledown_{l}^{2}\rho^{a^{\dagger},a}(\vec{n}):, \label{V_q2}%
\end{equation}%
\begin{equation}
V_{I^{2},q^{2}}=-\frac{1}{2}C_{I^{2},q^{2}}:\sum\limits_{\vec{n},I,l}\rho
_{I}^{a^{\dagger},a}(\vec{n})\triangledown_{l}^{2}\rho_{I}^{a^{\dagger}%
,a}(\vec{n}):, \label{V_I2q2}%
\end{equation}%
\begin{equation}
V_{S^{2},q^{2}}=-\frac{1}{2}C_{S^{2},q^{2}}:\sum\limits_{\vec{n},S,l}\rho
_{S}^{a^{\dagger},a}(\vec{n})\triangledown_{l}^{2}\rho_{S}^{a^{\dagger}%
,a}(\vec{n}):, \label{V_S2q2}%
\end{equation}%
\begin{equation}
V_{S^{2},I^{2},q^{2}}=-\frac{1}{2}C_{S^{2},I^{2},q^{2}}:\sum\limits_{\vec
{n},S,I,l}\rho_{S,I}^{a^{\dagger},a}(\vec{n})\triangledown_{l}^{2}\rho
_{S,I}^{a^{\dagger},a}(\vec{n}):, \label{V_S2I2q2}%
\end{equation}%
\begin{equation}
V_{(q\cdot S)^{2}}=\frac{1}{2}C_{(q\cdot S)^{2}}:\sum\limits_{\vec{n}}%
\sum\limits_{S}\Delta_{S}\rho_{S}^{a^{\dagger},a}(\vec{n})\sum
\limits_{S^{\prime}}\Delta_{S^{\prime}}\rho_{S^{\prime}}^{a^{\dagger},a}%
(\vec{n}):, \label{V_qs2}%
\end{equation}%
\begin{equation}
V_{I^{2},(q\cdot S)^{2}}=\frac{1}{2}C_{I^{2},(q\cdot S)^{2}}:\sum
\limits_{\vec{n},I}\sum\limits_{S}\Delta_{S}\rho_{S,I}^{a^{\dagger},a}(\vec
{n})\sum\limits_{S^{\prime}}\Delta_{S^{\prime}}\rho_{S^{\prime},I}%
^{a^{\dagger},a}(\vec{n}):, \label{V_I2qs2}%
\end{equation}%
\begin{equation}
V_{(iq\times S)\cdot k}=-\frac{i}{2}C_{(iq\times S)\cdot k}:\sum
\limits_{\vec{n},l,S,l^{\prime}}\varepsilon_{l,S,l^{\prime}}\left[  \Pi
_{l}^{a^{\dagger},a}(\vec{n})\Delta_{l^{\prime}}\rho_{S}^{a^{\dagger},a}%
(\vec{n})+\Pi_{l,S}^{a^{\dagger},a}(\vec{n})\Delta_{l^{\prime}}\rho
^{a^{\dagger},a}(\vec{n})\right]  :.
\end{equation}
These operators are different from those shown in Eq.~(\ref{V2}) and allow for
a simple projection onto different isospin channels. \ This will be useful
later when restricting to the interactions of neutrons.

The $V_{(iq\times S)\cdot k}$ term corresponds with the continuum interaction%
\begin{equation}
C_{(iq\times S)\cdot k}\left(  i\vec{q}\times\left(  \vec{\sigma}^{A}%
+\vec{\sigma}^{B}\right)  \right)  \cdot\vec{k},
\end{equation}
which vanishes unless the total spin is $S=1$. \ The continuum limit of this
interaction is antisymmetric under the exchange of $\vec{q}$ and $\vec{k}$ and
is nonzero only for odd parity channels. \ However the lattice interaction
$V_{(iq\times S)\cdot k}$ does not share this exact $t$-$u$ channel
antisymmetry at nonzero lattice spacing. \ Therefore $V_{(iq\times S)\cdot k}$
has small lattice artifacts for $S=1$ in even parity channels. \ We remove
this defect by including an explicit projection onto total isospin $I=1$,%
\begin{align}
V_{(iq\times S)\cdot k}^{I=1}  &  =-\frac{i}{2}C_{(iq\times S)\cdot k}%
^{I=1}\left\{  \frac{3}{4}:\sum\limits_{\vec{n},l,S,l^{\prime}}\varepsilon
_{l,S,l^{\prime}}\left[  \Pi_{l}^{a^{\dagger},a}(\vec{n})\Delta_{l^{\prime}%
}\rho_{S}^{a^{\dagger},a}(\vec{n})+\Pi_{l,S}^{a^{\dagger},a}(\vec{n}%
)\Delta_{l^{\prime}}\rho^{a^{\dagger},a}(\vec{n})\right]  :\right. \nonumber\\
&  \qquad+\left.  \frac{1}{4}:\sum\limits_{\vec{n},l,S,l^{\prime}%
,I}\varepsilon_{l,S,l^{\prime}}\left[  \Pi_{l,I}^{a^{\dagger},a}(\vec
{n})\Delta_{l^{\prime}}\rho_{S,I}^{a^{\dagger},a}(\vec{n})+\Pi_{l,S,I}%
^{a^{\dagger},a}(\vec{n})\Delta_{l^{\prime}}\rho_{I}^{a^{\dagger},a}(\vec
{n})\right]  :\right\}  .
\end{align}
This projection completely eliminates lattice artifacts in the $S=1$ even
parity channels.

\section{Scattering results for LO$_{3}$ and NLO$_{3}$}

We measure phase shifts and mixing angles using the spherical wall method
\cite{Borasoy:2007vy}. \ This consists of imposing a hard spherical wall
boundary on the relative separation between the two nucleons at some chosen
radius $R_{\text{wall}}$. \ Scattering phase shifts are determined from the
energies of the spherical standing waves, and mixing angles are extracted from
projections on to spherical harmonics. \ At next-to-leading order\ there are
nine unknown operator coefficients: $\ \Delta C$, $\Delta C_{I^{2}},$
$C_{q^{2}},$ $C_{I^{2},q^{2}},$ $C_{S^{2},q^{2}}$, $C_{S^{2},I^{2},q^{2}}$,
$C_{(q\cdot S)^{2}}$, $C_{I^{2},(q\cdot S)^{2}}$, and $C_{(iq\times S)\cdot
k}^{I=1}$. \ These nine operator coefficients are fit in the same manner as
described in Ref.~\cite{Borasoy:2007vi}. \ For $R_{\text{wall}}=10+\epsilon$
lattice units, where $\epsilon$ is an infinitesimal positive number, we
compute energy levels for the eight spherical wall modes listed in Table
\ref{fitvalues}. \ The labelling of these modes is discussed in
Ref.~\cite{Borasoy:2007vi}. \ In addition to these we also consider $Q_{d}$,
the quadrupole moment of the deuteron. $Q_{d}$\ is a measure of $S$-$D$
partial wave mixing at low energies and is somewhat easier to compute on the
lattice than\ the $S$-$D$ mixing angle. \begin{table}[tb]
\caption{Results for LO$_{3}$ and the physical target values}%
\label{fitvalues}
\begin{tabular}
[c]{||c|c|c|c||}\hline\hline
Spherical wave & Free nucleons & LO$_{3}$ & PWA93\\\hline
$1^{1}S_{0}$ (MeV) & $0.928$ & $0.418$ & $0.407$\\\hline
$3^{1}S_{0}$ (MeV) & $8.535$ & $6.843$ & $6.815$\\\hline
$1^{3}S(D)_{1}$ (MeV) & $0.928$ & $-2.225$ & $-2.225$\\\hline
$3^{3}S(D)_{1}$ (MeV) & $8.535$ & $5.430$ & $5.675$\\\hline
$2^{1}P_{1}$ (MeV) & $5.691$ & $5.755$ & $5.782$\\\hline
$2^{3}P(F)_{0}$ (MeV) & $5.691$ & $5.569$ & $5.584$\\\hline
$2^{3}P(F)_{1}$ (MeV) & $5.691$ & $5.754$ & $5.753$\\\hline
$2^{3}P(F)_{2}$ (MeV) & $5.691$ & $5.684$ & $5.669$\\\hline
$Q_{d}$ (fm$^{2}$) & N/A & $0.276$ & $0.286$\\\hline\hline
\end{tabular}
\end{table}For each of the nine observables we compute derivatives with
respect to the nine NLO coefficient operators. \ By inverting the resulting
$9\times9$ Jacobian matrix, we find values for the NLO coefficients needed to
match each of the nine target values using first-order perturbation theory.
\ The results for the operator coefficients are shown in Table
\ref{coefficients}.\begin{table}[tbtb]
\caption{Results for NLO$_{3}$ operator coefficients}%
\label{coefficients}
\begin{tabular}
[c]{||c|c||}\hline\hline
Coefficient & NLO$_{3}$\\\hline
$\Delta C$ (MeV$^{-2}$) & $-1.02\times10^{-5}$\\\hline
$\Delta C_{I^{2}}$ (MeV$^{-2}$) & $1.03\times10^{-5}$\\\hline
$C_{q^{2}}$ (MeV$^{-4}$) & $2.39\times10^{-10}$\\\hline
$C_{I^{2},q^{2}}$ (MeV$^{-4}$) & $-4.80\times10^{-11}$\\\hline
$C_{S^{2},q^{2}}$ (MeV$^{-4}$) & $1.67\times10^{-10}$\\\hline
$C_{S^{2},I^{2},q^{2}}$ (MeV$^{-4}$) & $-1.03\times10^{-10}$\\\hline
$C_{(q\cdot S)^{2}}$ (MeV$^{-4}$) & $-1.43\times10^{-10}$\\\hline
$C_{I^{2},(q\cdot S)^{2}}$ (MeV$^{-4}$) & $1.80\times10^{-10}$\\\hline
$C_{(iq\times S)\cdot k}^{I=1}$ (MeV$^{-4}$) & $1.60\times10^{-10}%
$\\\hline\hline
\end{tabular}
\end{table}

\begin{figure}
[ptb]
\begin{center}
\includegraphics[
height=2.6057in,
width=4.1943in
]%
{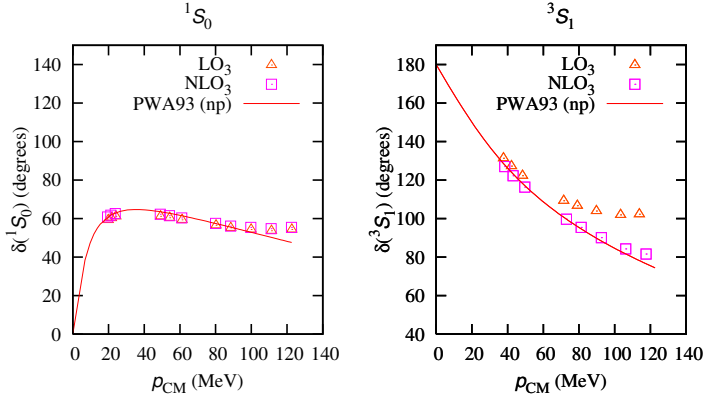}%
\caption{$S$-wave phase shifts versus center-of-mass momentum for LO$_{3}$ and
NLO$_{3}$.}%
\label{swave_b6swave_b0pwave}%
\end{center}
\end{figure}

With the NLO$_{3}$ coefficients in hand, we can now calculate lattice phase
shifts and mixing angles up to next-to-leading order using the spherical wall
method. \ We consider spherical walls with radii $R_{\text{wall}}=10+\epsilon
$, $9+\epsilon$, and $8+\epsilon$ lattice units. \ In order of increasing
momentum, the lattice data correspond with the first radial excitation for
$R_{\text{wall}}=10+\epsilon,9+\epsilon,$ and $8+\epsilon$; second radial
excitation of $R_{\text{wall}}=10+\epsilon,9+\epsilon,$ and $8+\epsilon;$ and
so on. \ The $S$-wave phase shifts for LO$_{3}$ and NLO$_{3}$ versus
center-of-mass momentum $p_{\text{CM}}$ are shown in
Fig.~\ref{swave_b6swave_b0pwave}. \ The NLO$_{3}$ results are in good
agreement with partial wave results from Ref.~\cite{Stoks:1993tb}.%

\begin{figure}
[ptbptb]
\begin{center}
\includegraphics[
height=2.527in,
width=2.2745in
]%
{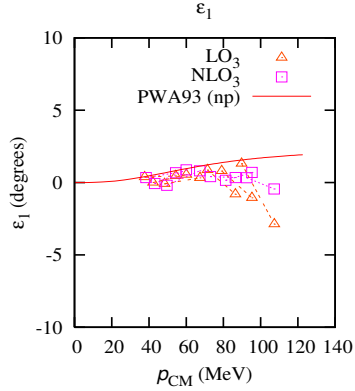}%
\caption{$\varepsilon_{1}$ mixing angle versus center-of-mass momentum for
LO$_{3}$ and NLO$_{3}.$}%
\label{eps1_b6swave_b0pwave}%
\end{center}
\end{figure}

We plot the $S$-$D$ mixing parameter $\varepsilon_{1}$ in the Stapp
parameterization \cite{Stapp:1956mz} in Fig.~\ref{eps1_b6swave_b0pwave}. \ The
pairs of points connected by dotted lines indicate pairs of coupled solutions
in the spherical wall formalism. \ While there are some deviations from the
partial wave data from Ref.~\cite{Stoks:1993tb}, the discrepancy is consistent
with effects produced by higher-order interactions.%
As expected the $S$-wave results for LO$_{3}$ are identical with LO$_{2}$
results in Ref.~\cite{Borasoy:2007vy}. \ In fact they agree in all even-$L$
partial wave channels. \ Results for NLO$_{3}$ and NLO$_{2}$ are also close
though not exactly the same. \ There are very small differences between the
two due to NLO interactions which are not completely separable into $S$-wave
and $P$-wave terms at nonzero lattice spacing.

\begin{figure}
[ptb]
\begin{center}
\includegraphics[
height=5.0289in,
width=4.1935in
]%
{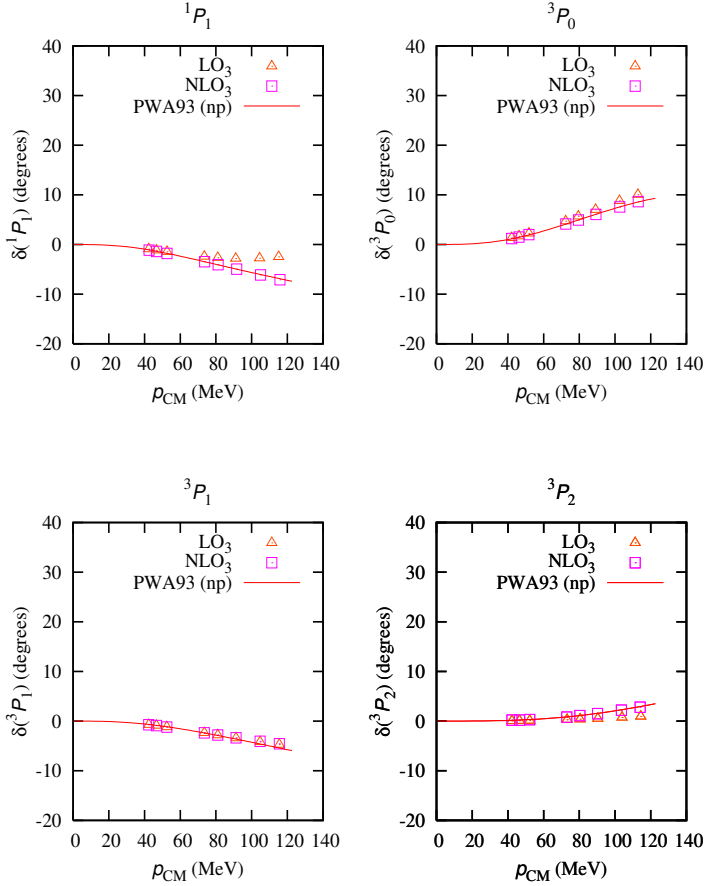}%
\caption{$P$-wave phase shifts versus center-of-mass momentum for LO$_{3}$ and
NLO$_{3}.$}%
\label{pwave_b6swave_b0pwave}%
\end{center}
\end{figure}

The $P$-wave phase shifts are shown in Fig.~\ref{pwave_b6swave_b0pwave}. \ We
see that the NLO$_{3}$ results match the partial wave data quite accurately.
\ Just as LO$_{3}$ and LO$_{2}$ agree in all even-$L$ partial wave channels,
LO$_{3}$ and LO$_{1}$ agree in all odd-$L$ partial wave channels. \ Results
for NLO$_{3}$ and NLO$_{1}$ are nearly identical, with only small differences
due to the numerical fitting of NLO coefficients on the lattice.%

\section{Auxiliary-field formalism for neutron matter}

So far we have been discussing general systems of low-energy nucleons with
both protons and neutrons. \ For computational efficiency we now specialize to
the case where all nucleons are neutrons. \ In this case all nucleon-nucleon
interactions are in the isospin-triplet channel. \ In the leading-order
transfer matrix $M_{\text{LO}_{3}}$ we can drop the
spin-triplet/isospin-singlet term involving $V_{S=1,I=0}(\vec{q})$ and make
the simplifying replacements,%
\begin{equation}
V_{S=0,I=1}(\vec{q})\rightarrow\frac{1}{8}:\rho^{a^{\dag},a}(\vec{q}%
)\rho^{a^{\dag},a}(-\vec{q}):-\frac{1}{8}:\sum_{S}\rho_{S}^{a^{\dag},a}%
(\vec{q})\rho_{S}^{a^{\dag},a}(-\vec{q}):,
\end{equation}%
\begin{equation}
\sum_{\substack{S_{1},S_{2},I}}\sum_{\vec{n}_{1},\vec{n}_{2}}G_{S_{1}S_{2}%
}(\vec{n}_{1}-\vec{n}_{2})\rho_{S_{1},I}^{a^{\dag},a}(\vec{n}_{1})\rho
_{S_{2},I}^{a^{\dag},a}(\vec{n}_{2})\rightarrow\sum_{S_{1},S_{2}}\sum_{\vec
{n}_{1},\vec{n}_{2}}G_{S_{1}S_{2}}(\vec{n}_{1}-\vec{n}_{2})\rho_{S_{1}%
}^{a^{\dag},a}(\vec{n}_{1})\rho_{S_{2}}^{a^{\dag},a}(\vec{n}_{2}).
\end{equation}
These modifications do not affect the interactions between neutrons and yields
the simplified transfer matrix,%
\begin{align}
M_{\text{LO}_{3}}  &  \rightarrow\colon\exp\left\{  -H_{\text{free}}\alpha
_{t}-\frac{C_{S=0,I=1}\alpha_{t}}{8L^{3}}\sum_{\vec{q}}f(\vec{q})\left[
\rho^{a^{\dag},a}(\vec{q})\rho^{a^{\dag},a}(-\vec{q})-\sum_{S}\rho
_{S}^{a^{\dag},a}(\vec{q})\rho_{S}^{a^{\dag},a}(-\vec{q})\right]  \right.
\nonumber\\
&  \qquad+\left.  \frac{g_{A}^{2}\alpha_{t}^{2}}{8f_{\pi}^{2}q_{\pi}}%
\sum_{S_{1},S_{2}}\sum_{\vec{n}_{1},\vec{n}_{2}}G_{S_{1}S_{2}}(\vec{n}%
_{1}-\vec{n}_{2})\rho_{S_{1}}^{a^{\dag},a}(\vec{n}_{1})\rho_{S_{2}}^{a^{\dag
},a}(\vec{n}_{2})\right\}  \colon. \label{MLO}%
\end{align}
At next-to-leading order the simplified neutron transfer matrix is%
\begin{align}
M_{\text{NLO}_{3}}  &  \rightarrow\colon\exp\left\{  -H_{\text{free}}%
\alpha_{t}-\frac{C_{S=0,I=1}\alpha_{t}}{8L^{3}}\sum_{\vec{q}}f(\vec{q})\left[
\rho^{a^{\dag},a}(\vec{q})\rho^{a^{\dag},a}(-\vec{q})-\sum_{S}\rho
_{S}^{a^{\dag},a}(\vec{q})\rho_{S}^{a^{\dag},a}(-\vec{q})\right]  \right.
\nonumber\\
&  -\left.  \alpha_{t}\left[  \frac{\Delta C^{I=1}}{\Delta C}\Delta
V+\frac{C_{q^{2}}^{I=1}}{C_{q^{2}}}V_{q^{2}}+\frac{C_{S^{2},q^{2}}^{I=1}%
}{C_{S^{2},q^{2}}}V_{S^{2},q^{2}}+\frac{C_{(q\cdot S)^{2}}^{I=1}}{C_{(q\cdot
S)^{2}}}V_{(q\cdot S)^{2}}+V_{(iq\times S)\cdot k}^{I=1}\right]  \right.
\nonumber\\
&  +\left.  \frac{g_{A}^{2}\alpha_{t}^{2}}{8f_{\pi}^{2}q_{\pi}}\sum
_{S_{1},S_{2},}\sum_{\vec{n}_{1},\vec{n}_{2}}G_{S_{1}S_{2}}(\vec{n}_{1}%
-\vec{n}_{2})\rho_{S_{1}}^{a^{\dag},a}(\vec{n}_{1})\rho_{S_{2}}^{a^{\dag}%
,a}(\vec{n}_{2})\right\}  \colon. \label{MNLO}%
\end{align}
In the following we use these simplified forms for the leading-order and
next-to-leading-order transfer matrices.

The transfer matrices in Eq.~(\ref{MLO}) and (\ref{MNLO}) can be rewritten in
terms of one-body interactions with auxiliary fields. \ The exact equivalence
between lattice formalisms with and without auxiliary fields is detailed in
\cite{Lee:2006hr,Borasoy:2006qn,Lee:2008fa}. \ We summarize the results here.

In neutron-neutron scattering only the neutral pion contributes to one-pion
exchange. \ We have been writing the rescaled neutral pion field as $\pi
_{3}^{\prime}$, but now we drop the subscript \textquotedblleft%
3\textquotedblright\ and simply write $\pi^{\prime}$. \ Let $M^{(n_{t})}%
(\pi^{\prime},s,s_{S})$ be the leading-order\ auxiliary-field transfer matrix
at time step $n_{t}$,%
\begin{align}
M^{(n_{t})}(\pi^{\prime},s,s_{S})  &  =\colon\exp\left\{  -H_{\text{free}%
}\alpha_{t}+\frac{g_{A}\alpha_{t}}{2f_{\pi}\sqrt{q_{\pi}}}%
{\displaystyle\sum_{\vec{n},S}}
\Delta_{S}\pi^{\prime}(\vec{n},n_{t})\rho_{S}^{a^{\dag},a}(\vec{n})\right.
\nonumber\\
&  \qquad\qquad\left.  +\frac{1}{2}\sqrt{-C_{S=0,I=1}\alpha_{t}}\sum_{\vec{n}%
}s(\vec{n},n_{t})\rho^{a^{\dag},a}(\vec{n})\right. \nonumber\\
&  \qquad\qquad\left.  +\frac{i}{2}\sqrt{-C_{S=0,I=1}\alpha_{t}}\sum_{\vec
{n},S}s_{S}(\vec{n},n_{t})\rho_{S}^{a^{\dag},a}(\vec{n})\right\}  \colon.
\end{align}
We can write $M_{\text{LO}_{3}}$ as the normalized integral%
\begin{equation}
M_{\text{LO}_{3}}=\frac{%
{\displaystyle\int}
D\pi^{\prime}DsDs_{S}\;e^{-S_{\pi\pi}^{(n_{t})}-S_{ss}^{(n_{t})}}M^{(n_{t}%
)}(\pi^{\prime},s,s_{S})}{%
{\displaystyle\int}
D\pi^{\prime}DsDs_{S}\;e^{-S_{\pi\pi}^{(n_{t})}-S_{ss}^{(n_{t})}}},
\label{LOaux}%
\end{equation}
where $S_{\pi\pi}^{(n_{t})}$ is the piece of the instantaneous pion action\ in
Eq.~(\ref{pionaction}) containing the neutral pion field at time step $n_{t}$,%
\begin{equation}
S_{\pi\pi}^{(n_{t})}(\pi^{\prime})=\frac{1}{2}\sum_{\vec{n}}\pi^{\prime}%
(\vec{n},n_{t})\pi^{\prime}(\vec{n},n_{t})-\frac{\alpha_{t}}{q_{\pi}}%
\sum_{\vec{n},l}\pi^{\prime}(\vec{n},n_{t})\pi^{\prime}(\vec{n}+\hat{l}%
,n_{t}),
\end{equation}
and $S_{ss}^{(n_{t})}$ is the auxiliary-field action at time step $n_{t}$,%
\begin{equation}
S_{ss}^{(n_{t})}(s,s_{S})=\frac{1}{2}\sum_{\vec{n},\vec{n}^{\prime}}s(\vec
{n},n_{t})f^{-1}(\vec{n}-\vec{n}^{\prime})s(\vec{n}^{\prime},n_{t})+\frac
{1}{2}\sum_{\vec{n},\vec{n}^{\prime},S}s_{S}(\vec{n},n_{t})f^{-1}(\vec{n}%
-\vec{n}^{\prime})s_{S}(\vec{n}^{\prime},n_{t}),
\end{equation}
with
\begin{equation}
f^{-1}(\vec{n}-\vec{n}^{\prime})=\frac{1}{L^{3}}\sum_{\vec{q}}\frac{1}%
{f(\vec{q})}e^{-i\vec{q}\cdot(\vec{n}-\vec{n}^{\prime})}.
\end{equation}

The NLO interactions require some additional auxiliary fields. \ Let%
\begin{align}
U^{(n_{t})}(\varepsilon)  &  =\sum_{\vec{n}}\varepsilon_{\rho}(\vec{n}%
,n_{t})\rho^{a^{\dagger},a}(\vec{n})+\sum_{\vec{n},S}\varepsilon_{\rho_{S}%
}(\vec{n},n_{t})\rho_{S}^{a^{\dagger},a}(\vec{n})+\sum_{\vec{n},S}%
\varepsilon_{\Delta_{S}\rho}(\vec{n},n_{t})\Delta_{S}\rho^{a^{\dagger},a}%
(\vec{n})\nonumber\\
&  +\sum_{\vec{n},S,S^{\prime}}\varepsilon_{\Delta_{S}\rho_{S^{\prime}}}%
(\vec{n},n_{t})\Delta_{S}\rho_{S^{\prime}}^{a^{\dagger},a}(\vec{n})+\sum
_{\vec{n},l}\varepsilon_{\triangledown_{l}^{2}\rho}(\vec{n},n_{t}%
)\triangledown_{l}^{2}\rho^{a^{\dagger},a}(\vec{n})\nonumber\\
&  +\sum_{\vec{n},l,S}\varepsilon_{\triangledown_{l}^{2}\rho_{S}}(\vec
{n},n_{t})\triangledown_{l}^{2}\rho_{S}^{a^{\dagger},a}(\vec{n})+\sum_{\vec
{n},l}\varepsilon_{\Pi_{l}}(\vec{n},n_{t})\Pi_{l}^{a^{\dagger},a}(\vec
{n})+\sum_{\vec{n},l,S}\varepsilon_{\Pi_{l,S}}(\vec{n},n_{t})\Pi
_{l,S}^{a^{\dagger},a}(\vec{n}).
\end{align}
With these extra fields and linear functional $U^{(n_{t})}(\varepsilon)$ we
define%
\begin{align}
M^{(n_{t})}(\pi^{\prime},s,s_{S},\varepsilon)  &  =\colon\exp\left\{
-H_{\text{free}}\alpha_{t}+\frac{g_{A}\alpha_{t}}{2f_{\pi}\sqrt{q_{\pi}}}%
{\displaystyle\sum_{\vec{n},S}}
\Delta_{S}\pi^{\prime}(\vec{n},n_{t})\rho_{S}^{a^{\dag},a}(\vec{n})\right.
\nonumber\\
&  \qquad\qquad\left.  +\frac{1}{2}\sqrt{-C_{S=0,I=1}\alpha_{t}}\sum_{\vec{n}%
}s(\vec{n},n_{t})\rho^{a^{\dag},a}(\vec{n})\right. \nonumber\\
&  \qquad\qquad\left.  +\frac{i}{2}\sqrt{-C_{S=0,I=1}\alpha_{t}}\sum_{\vec
{n},S}s_{S}(\vec{n},n_{t})\rho_{S}^{a^{\dag},a}(\vec{n})+\sqrt{\alpha_{t}%
}U^{(n_{t})}(\varepsilon)\right\}  \colon.
\end{align}
We also define the normalized integral,%
\begin{equation}
M^{(n_{t})}(\varepsilon)=\frac{%
{\displaystyle\int}
D\pi^{\prime}DsDs_{S}\;e^{-S_{\pi\pi}^{(n_{t})}-S_{ss}^{(n_{t})}}M^{(n_{t}%
)}(\pi^{\prime},s,s_{S},\varepsilon)}{%
{\displaystyle\int}
D\pi^{\prime}DsDs_{S}\;e^{-S_{\pi\pi}^{(n_{t})}-S_{ss}^{(n_{t})}}}.
\label{LOaux_eps}%
\end{equation}
When all $\varepsilon$ fields are set to zero we recover $M_{\text{LO}_{3}}$,%
\begin{equation}
M^{(n_{t})}(0)=M_{\text{LO}_{3}}\text{.}%
\end{equation}
To first order the NLO interactions in $M_{\text{NLO}_{3}}$ can be written as
a sum of bilinear derivatives of $M^{(n_{t})}(\varepsilon)$ with respect to
the $\varepsilon$ fields at $\varepsilon=0$,%
\begin{align}
M_{\text{NLO}_{3}}  &  =M_{\text{LO}_{3}}\nonumber\\
&  -\frac{1}{2}\Delta C^{I=1}\sum_{\vec{n}}\left.  \frac{\delta}%
{\delta\varepsilon_{\rho}(\vec{n},n_{t})}\frac{\delta}{\delta\varepsilon
_{\rho}(\vec{n},n_{t})}M^{(n_{t})}(\varepsilon)\right\vert _{\varepsilon
=0}\nonumber\\
&  +\frac{1}{2}C_{q^{2}}^{I=1}\sum_{\vec{n}}\left.  \frac{\delta}%
{\delta\varepsilon_{\rho}(\vec{n},n_{t})}\frac{\delta}{\delta\varepsilon
_{\triangledown_{l}^{2}\rho}(\vec{n},n_{t})}M^{(n_{t})}(\varepsilon
)\right\vert _{\varepsilon=0}+\;\cdots.
\end{align}

\section{Euclidean-time projection Monte Carlo}

We extract the properties of the ground state using Euclidean-time projection.
\ We briefly summarize the calculation in continuous-time notation before
describing the transfer matrix calculation at nonzero temporal lattice
spacing. \ Let $\left\vert \Psi^{\text{free}}\right\rangle $ be a Slater
determinant of free-particle standing waves in a periodic cube for $N$
neutrons. \ Let $H_{\text{LO}_{3}}$ be the Hamiltonian at leading order and
$H_{\text{NLO}_{3}}$ be the Hamiltonian at next-to-leading order. \ Let
$H_{\text{SU(2)}\not \pi }$ be the same as $H_{\text{LO}_{3}}$, but with
one-pion exchange turned off by setting $g_{A}$ to zero. \ As the notation
suggests, $H_{\text{SU(2)}\not \pi }$ is invariant under an exact SU(2)
intrinsic-spin symmetry. \ We define a trial wavefunction%
\begin{equation}
\left\vert \Psi(t^{\prime})\right\rangle =\exp\left(  -H_{\text{SU}%
(2)\not \pi }t^{\prime}\right)  \left\vert \Psi^{\text{free}}\right\rangle .
\end{equation}
The operator $\exp\left(  -H_{\text{SU(2)}\not \pi }t^{\prime}\right)  $ acts
as an approximate low-energy filter. \ In the auxiliary-field Monte Carlo
calculation this part of the Euclidean-time propagation is positive definite
for any even number of neutrons invariant under the SU(2) intrinsic-spin
symmetry \cite{Lee:2004hc,Chen:2004rq,Lee:2007eu}. \ With this trial
wavefunction we define the amplitude,%
\begin{equation}
Z(t)=\left\langle \Psi(t^{\prime})\right\vert \exp\left(  -H_{\text{LO}_{3}%
}t\right)  \left\vert \Psi(t^{\prime})\right\rangle ,
\end{equation}
as well as the transient energy at Euclidean time $t$,%
\begin{equation}
E_{\text{LO}_{3}}(t)=-\frac{\partial}{\partial t}\left[  \ln Z(t)\right]  .
\end{equation}
In the limit of large $t$,%
\begin{equation}
\lim_{t\rightarrow\infty}E_{\text{LO}_{3}}(t)=E_{0,\text{LO}_{3}},
\end{equation}
where $E_{0,\text{LO}_{3}}$ is the energy of the lowest eigenstate $\left\vert
\Psi_{0}\right\rangle $ of $H_{\text{LO}_{3}}$ with nonzero inner product with
$\left\vert \Psi(t^{\prime})\right\rangle $.

To compute the expectation value of some general operator $O$ we define%
\begin{equation}
Z_{O}(t)=\left\langle \Psi(t^{\prime})\right\vert \exp\left(  -H_{\text{LO}%
_{3}}t/2\right)  O\,\exp\left(  -H_{\text{LO}_{3}}t/2\right)  \left\vert
\Psi(t^{\prime})\right\rangle .
\end{equation}
The expectation value of $O$ for $\left\vert \Psi_{0}\right\rangle $ is given
by the large $t$ limit,%
\begin{equation}
\lim_{t\rightarrow\infty}\frac{Z_{O}(t)}{Z(t)}=\left\langle \Psi
_{0}\right\vert O\left\vert \Psi_{0}\right\rangle .
\end{equation}
Corrections to the energy at next-to-leading order can be computed using
$O=H_{\text{NLO}_{3}}-H_{\text{LO}_{3}}$. \ In that case%
\begin{equation}
\lim_{t\rightarrow\infty}\frac{Z_{O}(t)}{Z(t)}=E_{0,\text{NLO}_{3}%
}-E_{0,\text{LO}_{3}},
\end{equation}
where $E_{0,\text{NLO}_{3}}$ is the ground state energy at next-to-leading order.

On the lattice we construct $\left\vert \Psi(t^{\prime})\right\rangle $ using%
\begin{equation}
\left\vert \Psi(t^{\prime})\right\rangle =\left(  M_{\text{SU(2)}\not \pi
}\right)  ^{L_{t_{o}}}\left\vert \Psi^{\text{free}}\right\rangle ,
\label{L_t_o}%
\end{equation}
where $t^{\prime}=L_{t_{o}}\alpha_{t}$ and $L_{t_{o}}$ is the number of
\textquotedblleft outer\textquotedblright\ time steps. \ The amplitude $Z(t)$
is defined as%
\begin{equation}
Z(t)=\left\langle \Psi(t^{\prime})\right\vert \left(  M_{\text{LO}_{3}%
}\right)  ^{L_{t_{i}}}\left\vert \Psi(t^{\prime})\right\rangle , \label{L_t_i}%
\end{equation}
where $t=L_{t_{i}}\alpha_{t}$ and $L_{t_{i}}$ is the number of
\textquotedblleft inner\textquotedblright\ time steps. \ The transient energy%
\begin{equation}
E_{\text{LO}_{3}}(t+\alpha_{t}/2)
\end{equation}
is given by the ratio of the amplitudes for $t$ and $t+\alpha_{t}$,%
\begin{equation}
e^{-E_{\text{LO}_{3}}(t+\alpha_{t}/2)\cdot\alpha_{t}}=\frac{Z(t+\alpha_{t}%
)}{Z(t)}.
\end{equation}
The ground state energy $E_{0,\text{LO}_{3}}$ equals the asymptotic limit,%
\begin{equation}
E_{0,\text{LO}_{3}}=\lim_{t\rightarrow\infty}E_{\text{LO}_{3}}(t+\alpha
_{t}/2).
\end{equation}

We calculate these Euclidean-time projection amplitudes using auxiliary
fields. \ For a given configuration of auxiliary and pion fields, the
contribution to the amplitude $Z(t)$ is proportional to the determinant of an
$N\times N$ matrix of one-body amplitudes, where $N$ is the number of
neutrons. \ Integrations over auxiliary and pion field configurations are
computed using hybrid Monte Carlo. \ Details of the method can be found in
Ref.~\cite{Lee:2005fk,Lee:2006hr,Borasoy:2006qn,Lee:2008fa}.

For the ground state energy at next-to-leading order we compute expectation
values of $M_{\text{NLO}_{3}}$ and $M_{\text{LO}_{3}}$ inserted in the middle
of a string of $M_{\text{LO}_{3}}$ transfer matrices,%
\begin{equation}
Z_{M_{\text{NLO}_{3}}}(t)=\left\langle \Psi(t^{\prime})\right\vert \left(
M_{\text{LO}_{3}}\right)  ^{L_{t_{i}}/2}M_{\text{NLO}_{3}}\left(
M_{\text{LO}_{3}}\right)  ^{L_{t_{i}}/2}\left\vert \Psi(t^{\prime
})\right\rangle ,
\end{equation}%
\begin{equation}
Z_{M_{\text{LO}_{3}}}(t)=\left\langle \Psi(t^{\prime})\right\vert \left(
M_{\text{LO}_{3}}\right)  ^{L_{t_{i}}/2}M_{\text{LO}_{3}}\left(
M_{\text{LO}_{3}}\right)  ^{L_{t_{i}}/2}\left\vert \Psi(t^{\prime
})\right\rangle .
\end{equation}
From the ratio of amplitudes,%
\begin{equation}
\frac{Z_{M_{\text{NLO}_{3}}}(t)}{Z_{M_{\text{LO}_{3}}}(t)}=1-\Delta
E_{\text{NLO}_{3}}(t)\alpha_{t}+\cdots,
\end{equation}
we define the transient NLO energy correction $\Delta E_{\text{NLO}_{3}}(t)$.
\ The ellipsis denotes terms which are beyond first order in the
NLO\ coefficients. \ The NLO ground state energy $E_{0,\text{NLO}_{3}}$ is
calculated using%
\begin{equation}
E_{0,\text{NLO}_{3}}=E_{0,\text{LO}_{3}}+\lim_{t\rightarrow\infty}\Delta
E_{0,\text{NLO}_{3}}(t).
\end{equation}

\section{Precision tests of Monte Carlo simulations}

We use the two-neutron system to test the auxiliary-field Monte Carlo
simulations. \ The same observables are calculated using both auxiliary-field
Monte Carlo and the exact transfer matrix without auxiliary fields. \ We
choose a small system so that stochastic errors are small enough to expose
disagreement at the $0.1\%-1\%$ level. \ We choose the spatial length of
lattice to be $L=4$ and set the outer time steps $L_{t_{o}}=2$ and inner time
steps $L_{t_{i}}=2$. \ With $16$ processors we generate a total of about
$8\times10^{5}$ hybrid Monte Carlo trajectories. \ Each processor runs
completely independent trajectories, and we compute averages and stochastic
errors by comparing the results of all processors.

For the first test we choose $\left\vert \Psi^{\text{free}}\right\rangle $ to
be a spin-singlet state built from the Slater determinant of standing waves
$\left\vert \psi_{1}\right\rangle $ and $\left\vert \psi_{2}\right\rangle $
with%
\begin{equation}
\left\langle 0\right\vert a_{i,j}(\vec{n})\left\vert \psi_{1}\right\rangle
\propto\delta_{i,0}\delta_{j,1},\qquad\left\langle 0\right\vert a_{i,j}%
(\vec{n})\left\vert \psi_{2}\right\rangle \propto\delta_{i,1}\delta_{j,1}.
\end{equation}
For the second test we choose a spin-triplet state with standing waves%
\begin{equation}
\left\langle 0\right\vert a_{i,j}(\vec{n})\left\vert \psi_{1}\right\rangle
\propto\delta_{i,0}\delta_{j,1}\cos\tfrac{2\pi n_{1}}{L},\qquad\left\langle
0\right\vert a_{i,j}(\vec{n})\left\vert \psi_{2}\right\rangle \propto
\delta_{i,0}\delta_{j,1}\sin\tfrac{2\pi n_{1}}{L}.
\end{equation}
Comparisons between Monte Carlo results (MC) and exact transfer matrix
calculations (exact) are shown in Table \ref{precision}. \ The numbers in
parentheses are the estimated stochastic errors. \ \begin{table}[tb]
\caption{Monte Carlo results versus exact transfer matrix calculations for the
two-neutron spin singlet $S=0$ and spin triplet $S=1.$}%
\label{precision}
\begin{tabular}
[c]{||c|c|c|c|c||}\hline\hline
& $S=0$ (MC) & $S=0$ (exact) & $S=1$ (MC) & $S=1$ (exact)\\\hline
$E_{\text{LO}_{3}}(t+\alpha_{t}/2)$ [MeV] & $-2.90(2)$ & $-2.9112$ & $28.3(2)$
& $28.3658$\\\hline
$\frac{\partial\left(  \Delta E_{\text{NLO}_{3}}(t)\right)  }{\partial\left(
\Delta C^{I=1}\right)  }$ [$10^{4}$ MeV$^{3}$] & $4.751(5)$ & $4.7487$ &
$0.0003(7)$ & $0$\\\hline
$\frac{\partial\left(  \Delta E_{\text{NLO}_{3}}(t)\right)  }{\partial\left(
C_{q^{2}}^{I=1}\right)  }$ [$10^{9}$ MeV$^{5}$] & $1.580(2)$ & $1.5789$ &
$-1.025(4)$ & $-1.0264$\\\hline
$\frac{\partial\left(  \Delta E_{\text{NLO}_{3}}(t)\right)  }{\partial\left(
C_{S^{2},q^{2}}^{I=1}\right)  }$ [$10^{9}$ MeV$^{5}$] & $-4.741(6)$ &
$-4.7366$ & $-1.023(6)$ & $-1.0264$\\\hline
$\frac{\partial\left(  \Delta E_{\text{NLO}_{3}}(t)\right)  }{\partial\left(
C_{(q\cdot S)^{2}}^{I=1}\right)  }$ [$10^{8}$ MeV$^{5}$] & $-5.788(8)$ &
$-5.7818$ & $2.51(2)$ & $2.5533$\\\hline
$\frac{\partial\left(  \Delta E_{\text{NLO}_{3}}(t)\right)  }{\partial\left(
C_{(iq\times S)\cdot k}^{I=1}\right)  }$ [$10^{7}$ MeV$^{5}$] & $0.018(13)$ &
$0$ & $3.27(16)$ & $3.4534$\\\hline
$\Delta E_{\text{NLO}_{3}}(t)$ [MeV] & $-0.01655(8)$ & $-0.016440$ &
$-0.2455(9)$ & $-0.24559$\\\hline\hline
\end{tabular}
\end{table}The agreement between Monte Carlo results and exact transfer
calculations is consistent with the estimated stochastic errors.

\section{Results}

We simulate the ground state for $N=8,12,16$ neutrons on periodic cube
lattices. \ For $N=8$ we consider cube lengths $L=4,5,6,7$ lattice units.
\ For $N=12$, we use $L=5,6,7$, and for $N=16$ we use $L=6,7$. \ For each
value of $N$ and $L$ we fix $L_{t_{o}}$ at either $8$ or $10$ and vary
$L_{t_{i}}$ from $2$ up to $12$. \ For $\left\vert \Psi^{\text{free}%
}\right\rangle $ we take the Slater determinant formed by standing waves%
\begin{equation}
\left\langle 0\right\vert a_{i,j}(\vec{n})\left\vert \psi_{2k+1}\right\rangle
\propto\delta_{i,0}\delta_{j,1}f_{k}(\vec{n}),\qquad\left\langle 0\right\vert
a_{i,j}(\vec{n})\left\vert \psi_{2k+2}\right\rangle \propto\delta_{i,1}%
\delta_{j,1}f_{k}(\vec{n}),
\end{equation}
where%
\begin{equation}
f_{0}(\vec{n})=1,\quad f_{1}(\vec{n})=\cos\tfrac{2\pi n_{3}}{L},\quad
f_{2}(\vec{n})=\sin\tfrac{2\pi n_{3}}{L},\quad f_{3}(\vec{n})=\cos\tfrac{2\pi
n_{1}}{L},
\end{equation}%
\begin{equation}
f_{4}(\vec{n})=\sin\tfrac{2\pi n_{1}}{L},\quad f_{5}(\vec{n})=\cos\tfrac{2\pi
n_{2}}{L},\quad f_{6}(\vec{n})=\sin\tfrac{2\pi n_{2}}{L},\quad f_{7}(\vec
{n})=\cos\tfrac{2\pi(n_{1}+n_{2})}{L}.
\end{equation}
For $N=8$ the values of $k$ span the range $0\leq k\leq3$. \ For $N=12$,
$0\leq k\leq5$, and for $N=16$, $0\leq k\leq7$. \ For each value of $L_{t_{i}%
}$ a total of about $5\times10^{6}$ hybrid Monte Carlo trajectories are
generated by $2048$ processors, each running completely independent
trajectories. \ Averages and stochastic errors are computed by comparing the
results of all processors.

\begin{figure}
[ptb]
\begin{center}
\includegraphics[
height=5.9326in,
width=4.0491in
]%
{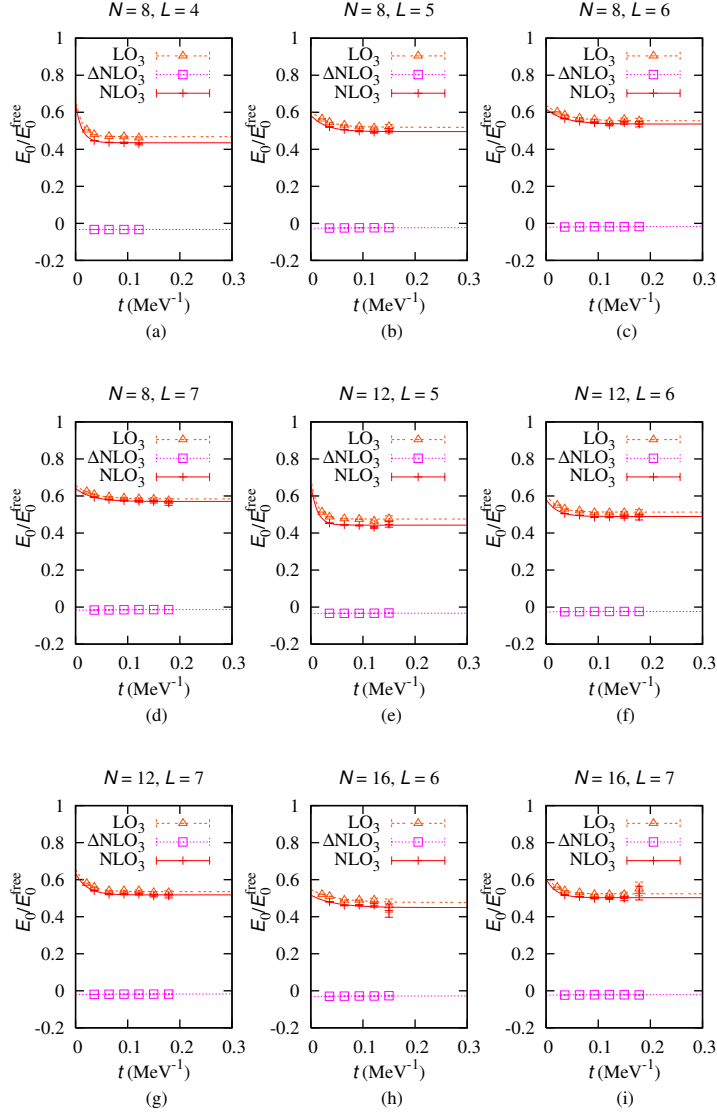}%
\caption{Plots of the three energy ratios defined in Eq.~(\ref{ratios}) versus
Euclidean projection time $t$. \ These are labelled as LO$_{3}$, $\Delta
$NLO$_{3}$, NLO$_{3}$ respectively.}%
\label{alln_alll_b6swave_b0pwave}%
\end{center}
\end{figure}

Let $E_{0}^{\text{free}}$ be the energy of the ground state for noninteracting
neutrons. \ In Fig.~\ref{alln_alll_b6swave_b0pwave} we show the dimensionless
ratios%
\begin{equation}
\frac{E_{\text{LO}_{3}}(t)}{E_{0}^{\text{free}}}\text{,\quad}\frac{\Delta
E_{\text{NLO}_{3}}(t)}{E_{0}^{\text{free}}},\text{\quad}\frac{E_{\text{LO}%
_{3}}(t)+\Delta E_{\text{NLO}_{3}}(t)}{E_{0}^{\text{free}}}\text{,}
\label{ratios}%
\end{equation}
versus Euclidean time $t$. \ These are labelled using the shorthand LO$_{3}$,
$\Delta$NLO$_{3}$, and NLO$_{3}$ respectively. \ In addition to the Monte
Carlo data we plot the asymptotic expressions,%
\begin{equation}
\frac{E_{\text{LO}_{3}}(t)}{E_{0}^{\text{free}}}\approx\frac{E_{0,\text{LO}%
_{3}}}{E_{0}^{\text{free}}}+Ae^{-\delta E\cdot t}, \label{asymptotic1}%
\end{equation}%
\begin{equation}
\frac{\Delta E_{\text{NLO}_{3}}(t)}{E_{0}^{\text{free}}}\approx\frac
{E_{0,\text{NLO}_{3}}-E_{0,\text{LO}_{3}}}{E_{0}^{\text{free}}}+Be^{-\delta
E\cdot t/2}. \label{asymptotic2}%
\end{equation}%
\begin{equation}
\frac{E_{\text{LO}_{3}}(t)+\Delta E_{\text{NLO}_{3}}(t)}{E_{0}^{\text{free}}%
}\approx\frac{E_{0,\text{NLO}_{3}}}{E_{0}^{\text{free}}}+Ae^{-\delta E\cdot
t}+Be^{-\delta E\cdot t/2}. \label{asymptotic3}%
\end{equation}
The unknown coefficients $A$ and $B$, energy gap $\delta E$, and ground state
energies $E_{0,\text{LO}_{3}}$ and $E_{0,\text{NLO}_{3}}$ are determined by
least squares fitting. \ The $e^{-\delta E\cdot t}$ dependence in
Eq.~(\ref{asymptotic1}) comes from the contribution of low-energy excitations
with energy gap $\delta E$ above the ground state. \ The $e^{-\delta E\cdot
t/2}$ dependence in Eq.~(\ref{asymptotic2}) comes from the matrix element of
$M_{\text{NLO}_{3}}$ between the ground state and low-energy excitations at
energy gap $\delta E$.%

The results of the asymptotic fits for $E_{0,\text{LO}_{3}}/E_{0}%
^{\text{free}}$ and $E_{0,\text{NLO}_{3}}/E_{0}^{\text{free}}$ are shown in
Table~\ref{fits}. \begin{table}[tb]
\caption{Fit results for $E_{0,\text{LO}_{3}}/E_{\text{0}}^{\text{free}}$ and
$E_{0,\text{NLO}_{3}}/E_{\text{0}}^{\text{free}}$.}%
\label{fits}%
\begin{tabular}
[c]{||c|c|c|c|c|c||}\hline\hline
$N$ & $L$ & $k_{F}$ (MeV) & $E_{0,\text{LO}_{3}}/E_{0}^{\text{free}}$ &
$E_{0,\text{NLO}_{3}}/E_{0}^{\text{free}}$ & $\chi^{2}$/d.f.\\\hline
$8$ & $4$ & $155$ & $0.469(2)$ & $0.436(2)$ & $0.6$\\\hline
$8$ & $5$ & $124$ & $0.519(4)$ & $0.496(4)$ & $0.8$\\\hline
$8$ & $6$ & $103$ & $0.554(4)$ & $0.537(4)$ & $0.8$\\\hline
$8$ & $7$ & $88$ & $0.584(8)$ & $0.571(8)$ & $0.5$\\\hline
$12$ & $5$ & $142$ & $0.476(2)$ & $0.443(2)$ & $1.8$\\\hline
$12$ & $6$ & $118$ & $0.513(2)$ & $0.490(2)$ & $2.0$\\\hline
$12$ & $7$ & $101$ & $0.535(3)$ & $0.518(3)$ & $1.0$\\\hline
$16$ & $6$ & $130$ & $0.477(10)$ & $0.450(10)$ & $1.4$\\\hline
$16$ & $7$ & $111$ & $0.524(3)$ & $0.503(3)$ & $2.0$\\\hline\hline
\end{tabular}
\end{table}On average the $\chi^{2}$ per degree of freedom for the fits is
around $1$. \ The error estimates for $E_{0,\text{LO}_{3}}/E_{0}^{\text{free}%
}$ and $E_{0,\text{NLO}_{3}}/E_{0}^{\text{free}}$ are calculated by explicit
simulation. \ We introduce Gaussian-random noise scaled by the error bars of
each data point. \ The fit is repeated many times with the random noise
included to estimate the one standard-deviation spread in the fit parameters.
\ In Table~\ref{fits} the Fermi momentum $k_{F}$ for each neutron spin is
calculated from the density of neutrons in the periodic cube,%
\begin{equation}
k_{F}=\frac{1}{L}\left(  3\pi^{2}N\right)  ^{1/3}.
\end{equation}

\section{Discussion}

\subsection{Comparisons with other results}

\begin{figure}
[ptb]
\begin{center}
\includegraphics[
height=3.026in,
width=3.5959in
]%
{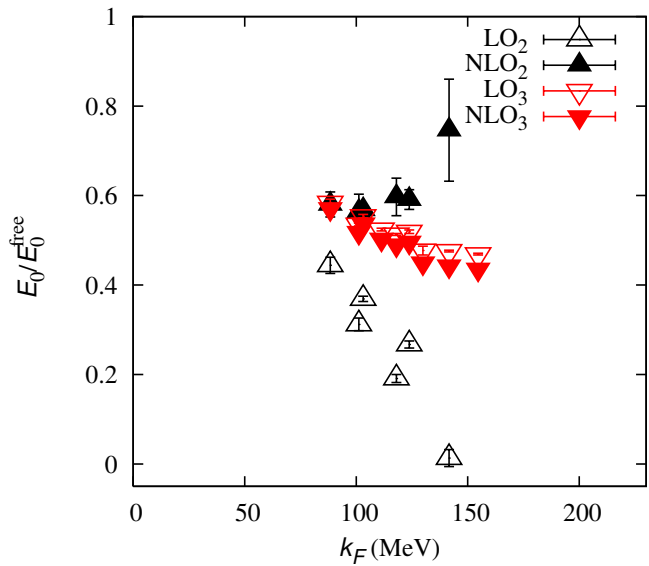}%
\caption{Comparison of the ground state energy ratio $E_{0}/E_{0}%
^{\text{free}}$ for LO$_{2}$, NLO$_{2}$, LO$_{3}$, and NLO$_{3}$ as a function
of $k_{F}$.}%
\label{kf_xsi_lo2_lo3}%
\end{center}
\end{figure}

In Fig.~\ref{kf_xsi_lo2_lo3} we compare the ground state energy ratio
$E_{0}/E_{0}^{\text{free}}$ for LO$_{2}$, NLO$_{2}$, LO$_{3}$, and NLO$_{3}$
as a function of $k_{F}$.%
We note two points here. \ First the difference between LO$_{3}$ and NLO$_{3}$
values for $E_{0}/E_{0}^{\text{free}}$ is relatively small over the range of
$k_{F}$ plotted. \ This suggests that the convergence of the effective field
theory expansion appears reliable, and the difference between LO$_{3}$ and
NLO$_{3}$ values provides an upper estimate on the size of contributions at
higher orders. \ Second the results for NLO$_{2}$ and NLO$_{3}$ agree for
$k_{F}$ less than $100$ MeV. \ This is the region where we expect the
perturbative treatment of NLO$_{2}$ corrections to be accurate. \ The
agreement with NLO$_{3}$ provides some confidence in the effective field
theory approach to dilute neutron matter. \ It is also an explicit test of
model independence at fixed lattice spacing as suggested in
Ref.~\cite{Borasoy:2007vi}.

\begin{figure}
[ptb]
\begin{center}
\includegraphics[
height=3.3589in,
width=3.7758in
]%
{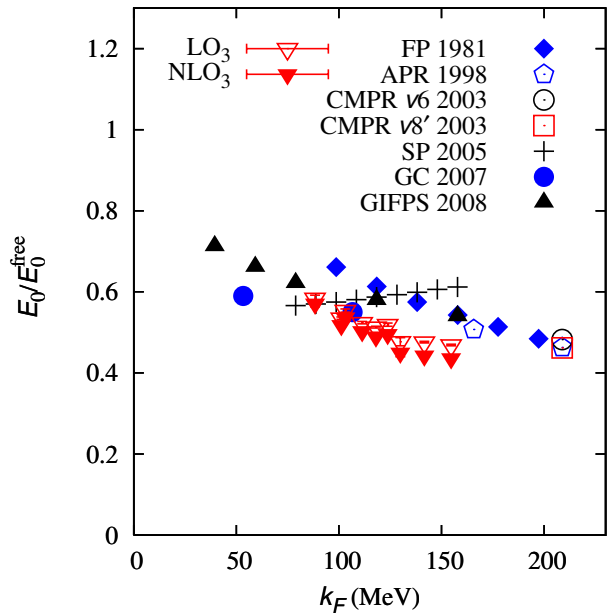}%
\caption{Ground state energy ratio $E_{0}/E_{0}^{\text{free}}$ for LO$_{3}$
and NLO$_{3}$ versus Fermi momentum $k_{F}$. \ For comparison we show results
for FP 1981 \cite{Friedman:1981qw}, APR 1998 \cite{Akmal:1998cf}, CMPR $v6$
and $v8^{\prime}$ 2003 \cite{Carlson:2003wm}, SP 2005 \cite{Schwenk:2005ka},
GC 2007 \cite{Gezerlis:2007fs}, and GIFPS 2008 \cite{Gandolfi:2008id}.}%
\label{kf_literature}%
\end{center}
\end{figure}

In Fig.~\ref{kf_literature} we compare ground state energies for LO$_{3}$ and
NLO$_{3}$ with other results from the literature: \ FP 1981
\cite{Friedman:1981qw}, APR 1998 \cite{Akmal:1998cf}, CMPR $v6$ and
$v8^{\prime}$ \cite{Carlson:2003wm}, SP 2005 \cite{Schwenk:2005ka}, GC 2007
\cite{Gezerlis:2007fs}, and GIFPS 2008 \cite{Gandolfi:2008id}. \ Compared with
other calculations our ground state energies are slightly lower for $k_{F}$
near $130$ MeV, but overall there is relatively good agreement.%

\subsection{Expansion near the unitarity limit}

The unitarity limit is an idealized limit of attractive two-component fermions
where the $S$-wave scattering length is infinite and the range of the
interaction is negligible. \ The $S$-wave scattering length for
neutron-neutron scattering is $-18.5$ fm, while the range of the interaction
is comparable to the Compton wavelength of the pion, $m_{\pi}^{-1}$. \ The
unitarity limit is approximately realized in neutron matter when the average
particle separation is between these two length scales. \ This occurs at a
Fermi momentum of about $80$ MeV. \ In the unitarity limit the ground state
has no dimensionful parameters other than particle density. \ Therefore the
ground state energy in the unitarity limit should obey a simple and universal
relation $E_{0}=\xi E_{0}^{\text{free}}$ for some dimensionless constant $\xi$.

The unitarity limit has been reproduced in trapped cold atom experiments using
$^{6}$Li and $^{40}$K. \ The scattering length is tuned to infinity using a
Feshbach resonance and the system is sufficiently dilute that the range of the
interaction is negligible. \ Recent experimental measurements for $\xi$ give
$0.32_{-13}^{+10}$ \cite{Bartenstein:2004}, $0.51(4)$ \cite{Kinast:2005},
$0.46_{-05}^{+12}$ \cite{Stewart:2006}, and $0.39(2)$ \cite{Luo:2008a}.
\ There have been numerous analytic calculations for $\xi$ varying over the
range from $0.2$ to $0.6$
\cite{Engelbrecht:1997,Baker:1999dg,Heiselberg:1999,Perali:2004,Schafer:2005kg,Papenbrock:2005,Nishida:2006a,Nishida:2006b,Arnold:2007,Nikolic:2007,Veillette:2006}%
. \ Several numerical calculations both on the lattice and in the continuum
find results varying from about $0.25$ to $0.45$
\cite{Carlson:2003z,Astrakharchik:2004,Bulgac:2005a,Burovski:2006a,Burovski:2006b,Lee:2005is,Lee:2005it,Lee:2007a,Abe:2007fe,Abe:2007ff,Juillet:2007a,Bulgac:2008c,Lee:2008xs}%
. \ The most recent of these numerical calculations agree on a smaller window
between $0.30$ and $0.40$.

For finite $S$-wave scattering length $a_{0}$ the deviation away from
unitarity can be parameterized as%
\begin{equation}
\frac{E_{0}}{E_{\text{0}}^{\text{free}}}\approx\xi-\frac{\xi_{1}}{k_{F}a_{0}}.
\end{equation}
\ As shorthand notation we define
\begin{equation}
f(k_{F}a_{0})=\xi-\frac{\xi_{1}}{k_{F}a_{0}}. \label{f}%
\end{equation}
In the recent literature there is general agreement on the value of $\xi_{1}$,
ranging from about $0.8$ to $1.0$
\cite{Chang:2004PRA,Astrakharchik:2004,Chen:2006A,Lee:2006hr,Abe:2007fe,Abe:2007ff,Lee:2007a}%
. \ In the following analysis we use the values $\xi=0.31$ and $\xi_{1}=0.81$
calculated in Ref.~\cite{Lee:2007a}.

In addition to the corrections at finite scattering length we expect
corrections proportional to $k_{F}r_{0}$ due to the $S$-wave effective range
$r_{0}$. \ For neutron-neutron scattering $r_{0}$ is $2.7$ fm. \ We also
expect higher-order corrections away from the unitarity limit arising from
higher powers of $1/(k_{F}a_{0})$ and $k_{F}r_{0}$, as well as other terms
associated with the $S$-wave shape parameter and triplet $P$-wave scattering
volumes. \ In general we can write%
\begin{equation}
E_{0}/E_{\text{0}}^{\text{free}}\approx f(k_{F}a_{0})+c_{1}k_{F}r_{0}%
+c_{2}k_{F}^{2}m_{\pi}^{-2}+c_{3}k_{F}^{3}m_{\pi}^{-3}+\cdots\text{.}
\label{c1c2c3}%
\end{equation}
Due to the relatively narrow spread of $k_{F}$ values considered in our
lattice simulations, it is difficult to constrain $c_{2}$ and $c_{3}$ and
higher coefficient powers. \ However we can constrain the parameter $c_{1}$. \ 

\begin{figure}
[ptb]
\begin{center}
\includegraphics[
height=3.3243in,
width=3.8017in
]%
{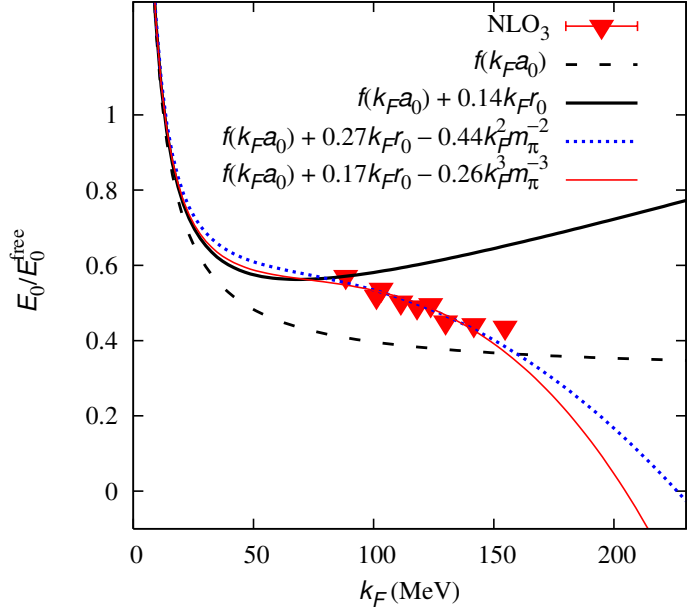}%
\caption{Comparison of $E_{0,\text{NLO}_{3}}/E_{\text{0}}^{\text{free}}$ with
various fits involving subsets of the unknown parameters $c_{1}$,$c_{2}%
$,$c_{3}$ as defined in Eq.~(\ref{c1c2c3}).}%
\label{kf_xsi_nlo3_unitarity}%
\end{center}
\end{figure}

If we set $c_{2}=c_{3}=0$ and determine $c_{1}$ from the data point with the
smallest value for $k_{F}$ we get%
\begin{equation}
E_{0,\text{NLO}_{3}}/E_{\text{0}}^{\text{free}}\approx f(k_{F}a_{0}%
)+0.14k_{F}r_{0}.
\end{equation}
If instead we set $c_{3}=0$ and determine $c_{1}$ and $c_{2}$ simultaneously
we find%
\begin{equation}
E_{0,\text{NLO}_{3}}/E_{\text{0}}^{\text{free}}\approx f(k_{F}a_{0}%
)+0.27k_{F}r_{0}-0.44k_{F}^{2}m_{\pi}^{-2}.
\end{equation}
As a third alternative if we set $c_{2}=0$ and fit $c_{1}$ and $c_{3}$
simultaneously, we get%
\begin{equation}
E_{0,\text{NLO}_{3}}/E_{\text{0}}^{\text{free}}\approx f(k_{F}a_{0}%
)+0.17k_{F}r_{0}-0.26k_{F}^{3}m_{\pi}^{-3}.
\end{equation}
The results of these fits are shown in Fig.~\ref{kf_xsi_nlo3_unitarity}. \ Our
simple analysis suggests a value for $c_{1}$ in the range between $0.14$ and
$0.27$. \ This is consistent with the value $0.15$ for the same coefficient
found in Ref.~\cite{Borasoy:2007vk}.%

\section{Summary}

We have presented lattice simulations for the ground state energy of dilute
neutron matter at next-to-leading order in chiral effective field theory. \ We
have solved some problems that arose in recent work using leading-order
lattice actions LO$_{1}$ and LO$_{2}$. \ LO$_{1}$ involved point-like
\textquotedblleft contact\textquotedblright\ interactions while LO$_{2}$ used
Gaussian-smeared \textquotedblleft contact\textquotedblright\ interactions.
\ In this work we introduced a new action LO$_{3}$ which equals LO$_{2}$ in
each $S$-wave channel and equals LO$_{1}$ in each $P$-wave channel. \ The
action was constructed using projection operators for the
spin-singlet/isospin-triplet and spin-triplet/isospin-singlet channels.
\ Using the spherical wall method we computed phase shifts and mixing angles
for the new lattice action up to next-to-leading order and fitted all unknown
operator coefficients.

In the auxiliary-field formalism we used Euclidean-time projection Monte Carlo
to compute the ground state energy of $N=8,12,16$ neutrons in a periodic cube,
covering a density range from 2\% to 10\% of normal nuclear density. \ For
$k_{F}$ less than $100$ MeV we found ground state energies at next-to-leading
order that agreed with earlier lattice results using the action NLO$_{2}$.
\ For $k_{F}$ greater than $100$ MeV we found that the new action leads to
much smaller corrections at next-to-leading order. \ The difference between
leading-order and next-to-leading-order values provides an upper estimate on
the size of contributions at higher orders. \ Though we find somewhat lower
values for the ground state energy near $k_{F}=130$ MeV, our results are in
general agreement with other calculations reported in the literature.

The ground state energy ratio $E_{0}/E_{\text{0}}^{\text{free}}$ was also
analyzed as an expansion about the unitarity limit. \ We considered
corrections due to finite scattering length, nonzero effective range, and
higher-order effects. \ If we use the parameterization%
\begin{equation}
E_{0}/E_{\text{0}}^{\text{free}}\approx f(k_{F}a_{0})+c_{1}k_{F}r_{0}%
+c_{2}k_{F}^{2}m_{\pi}^{-2}+c_{3}k_{F}^{3}m_{\pi}^{-3}+\cdots\text{,}%
\end{equation}
we find $c_{1}$ in the range from $0.14$ to $0.27$. \ In principle the
coefficient $c_{1}$ is a universal constant that can be measured in any
two-component fermionic system near the unitarity limit. \ Explicit tests of
this universality may be a subject for future investigation. \ With regard to
further investigations of neutron matter, future work on lattice simulations
should be considered at next-to-next-to-leading order in chiral effective
field theory. \ Simulations should also be done at smaller and larger lattice
spacings to check independence on the lattice spacing and to probe both higher
and lower densities.

\section*{Acknowledgements}

We are grateful for discussions with Bugra Borasoy. \ Partial financial
support from the Deutsche Forschungsgemeinschaft (SFB/TR 16), Helmholtz
Association (contract number VH-NG-222 and VH-VI-231), and U.S. Department of
Energy (DE-FG02-03ER41260) are acknowledged. \ This research is part of the EU
Integrated Infrastructure Initiative in Hadron Physics under contract number
RII3-CT-2004-506078. \ The computational resources for this project were
provided by the J\"{u}lich Supercomputing Centre at the Forschungszentrum J\"{u}lich.

\bibliographystyle{apsrev}
\bibliography{NuclearMatter}

\end{document}